%% file: main.tex
\documentclass[twocolumn ,amsmath,amssymb,aps,pra, floatfix]{revtex4-1}

\usepackage{graphicx}
\usepackage{physics}
\usepackage{comment}
\usepackage{color}
\usepackage{dcolumn}
\usepackage{bm}
\usepackage{svg}
\usepackage[colorlinks,urlcolor=blue,citecolor=blue,linkcolor=blue]{hyperref}
\usepackage{cleveref}

\newcommand{\up}{\uparrow}
\newcommand{\down}{\downarrow}

\renewcommand{\k}{{\bf k}}
\newcommand{\p}{{\bf p}}
\renewcommand{\P}{{\bf P}}
\newcommand{\q}{{\bf q}}
\newcommand{\Q}{{\bf Q}}

\newcommand{\eb}{\varepsilon_{X}}

\newcommand{\nn}{\nonumber}
\newcommand{\beq}{\begin{equation}}
\newcommand{\eeq}{\end{equation}}

\makeatletter
\renewcommand*\env@matrix[1][\arraystretch]{%
  \edef\arraystretch{#1}%
  \hskip -\arraycolsep
  \let\@ifnextchar\new@ifnextchar
  \array{*\c@MaxMatrixCols c}}
\makeatother

\newcommand{\dn}{\downarrow}

\usepackage{amsmath, url, graphicx,float, braket, simpler-wick, hyperref, amsfonts,amsthm,bm, bbm}
\usepackage{graphicx}
\usepackage{dcolumn}
\usepackage{bm}
\allowdisplaybreaks
\begin{document}

\title{Polaronic polariton quasiparticles in a dark excitonic medium}

\author{Kenneth Choo}
\author{Olivier Bleu}
\author{Jesper Levinsen}
\author{Meera M. Parish}
\affiliation{
School of Physics and Astronomy, Monash University, Victoria 3800, Australia \\
and ARC Centre of Excellence in Future Low-Energy Electronics Technologies, Monash University, Victoria 3800, Australia
}

\date{\today}

\begin{abstract}
Exciton polaritons are hybrid 
particles of excitons (bound electron-hole pairs) and cavity photons, which are renowned for displaying Bose Einstein condensation and other coherent phenomena at elevated temperatures. However, their formation in semiconductor microcavities is often accompanied by the appearance of an incoherent bath of optically dark excitonic states that can interact with polaritons via their matter component. Here we show that the presence of such a dark excitonic medium can ``dress'' polaritons with density fluctuations to form coherent polaron-like quasiparticles, thus fundamentally modifying their character. We employ a many-body Green's function approach that naturally incorporates correlations beyond the standard mean-field theories applied to this system. With increasing exciton density, we find a reduction in the light-matter coupling that arises from the polaronic dressing cloud rather than any saturation induced by the fermionic constituents of the exciton. In particular, we observe the strongest effects when the spin of the polaritons is opposite that of the excitonic medium. In this case, the coupling to light generates an additional polaron quasiparticle---the biexciton polariton---which emerges due to the dark-exciton counterpart of a polariton Feshbach resonance. Our results can explain recent experiments on polariton interactions in two-dimensional semiconductors and potentially provide a route to tailoring the properties of exciton polaritons and their correlations.
\end{abstract}

\maketitle

\section{Introduction}
The concept of quasiparticles has revolutionized our understanding of complex quantum systems. A famous example is Landau's quasiparticles within Fermi liquid theory~\cite{Abrikosov1959,PinesNozieres1966} which describe the collective behavior of interacting fermions, and which have provided a remarkably successful description of Fermi systems ranging from liquid $^3$He to the semiconductors that underpin modern electronics. Another important example is the so-called \emph{polaron} quasiparticle, which forms when a mobile impurity particle becomes ``dressed'' by excitations of a background quantum medium. This latter scenario 
was initially confined to the case of electrons interacting with phonons in a crystal lattice \cite{Devreese_2009} 
but has since emerged in a variety of systems, including ultracold atomic gases \cite{Massignan2014,Scazza2022review} and, most recently, doped semiconductors~\cite{sidler2017,Efimkin2017}. 

\begin{figure}[htb]
\includegraphics[width=1\columnwidth]{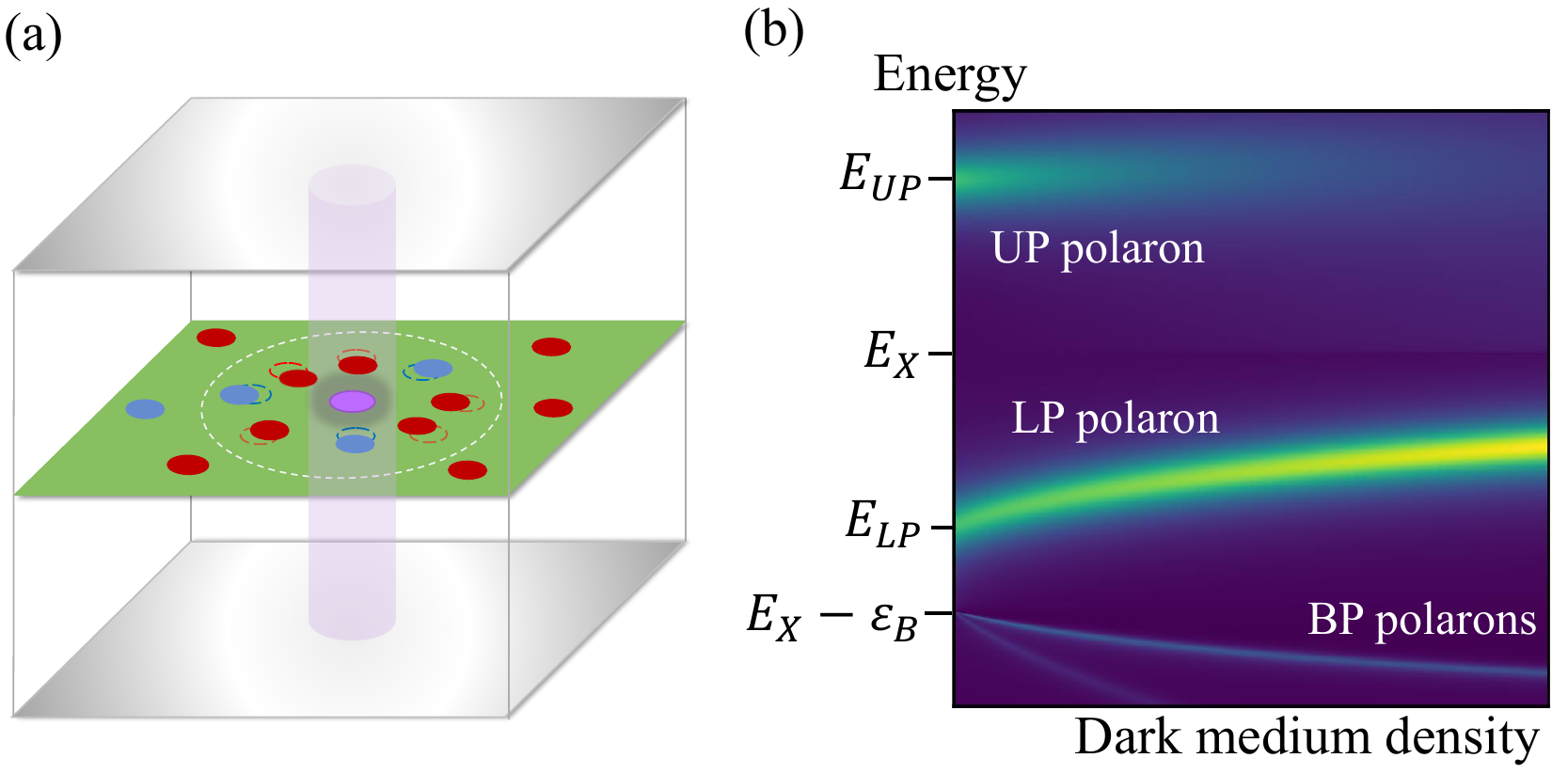}
\caption{(a) Schematic of a polaron-like quasiparticle in a semiconductor optical microcavity. An exciton polariton (purple circle and region) interacts with a dark excitonic medium in a 2D semiconductor (green layer) which may have two different spins (red and blue circles). Depending on the spin and photon energy, the polariton either attracts or repels the excitons, resulting in an increase or decrease in the local exciton density, respectively. 
(b) A typical energy spectrum at zero photon-exciton detuning that shows how the upper (UP) and lower (LP) polaritons at energies $E_{UP}, E_{LP}$ evolve into dressed polarons with increasing dark medium density. Below the exciton energy $E_X$, an additional polaron quasiparticle emerges that is connected to the biexciton with binding energy $\varepsilon_{B}$ at vanishing density. This biexciton polariton (BP) is sensitive to the spin composition of the medium and can exhibit significant Zeeman splitting.}
\label{fig:overallsetup}
\end{figure}

In this work, we extend the quasiparticle paradigm, as embodied by polarons and Landau's quasiparticles, and show that it provides a new framework for understanding hybrid light-matter systems in semiconductor microcavities, as depicted in Fig.~\ref{fig:overallsetup}.
Here cavity photons are strongly coupled to excitons---bound electron-hole pairs---in a two-dimensional (2D) semiconductor, leading  to the formation of exciton polaritons (or polaritons), 
emergent hybrid particles that are superpositions of light and matter~\cite{Weisbuch1992,kavokin2003cavity,deng2010,carusotto2013review}. 
Most notably, polaritons possess a small mass inherited from their photonic component, allowing them to Bose condense at high temperatures~\cite{byrnes2014}, while their excitonic component can potentially generate strong optical nonlinearities~\cite{ferrier2011,delteil2019,MunozMatutano2019}.

The polaritons' hybrid nature makes this platform attractive for the investigation of a rich variety of phenomena ranging from superfluidity \cite{amo2009superfluidity,lagoudakis2008quantized,Sanvitto2010, Lerario2017} and topological effects \cite{StJean2017,Klembt2018,Gianfrate2020,Solnyshkov2021} to the emulation of the classical XY and Kardar-Parisi-Zhang models \cite{Berloff2017,Fontaine2021}. Polaritons also have potential applications in optoelectronics~\cite{ballarini2013, nguyen2013, marsault2015, zasedatelev2019,Sanvitto2016}, while the recent advent of atomically thin materials with robust exciton bound states further widens the capabilities of light-matter systems~\cite{WangRMP2018,schneider2018review}. On the other hand, it is well known that 
such microcavity polaritons are often accompanied by a ``dark reservoir''~\cite{deng2010,carusotto2013review} consisting of excitons  which are not coupled to light (e.g., momentum forbidden) yet can interact with polaritons via their excitonic part and thus impact the behavior of polaritons. Furthermore, the density of dark excitons can be substantial~\cite{Madeo2020,pieczarka2022}, even when polaritons are directly excited with a resonant laser~\cite{gavrilov2012,Vishnevsky2012,Takemura2016,Walker2017,stepanov2019}.

We show here that the presence of a dark excitonic medium can fundamentally modify the polaritons themselves by dressing them to form polaron-like quasiparticles, as illustrated in Fig.~\ref{fig:overallsetup}(a). 
In addition to shifting the energy of the polaritons, the polaronic dressing cloud can reduce the polariton's light-matter coupling and lifetime, as well as generate a new quasiparticle---the biexciton polariton---a polaron that is adiabatically connected to the biexciton (two bound excitons) at vanishing exciton density  [Fig.~\ref{fig:overallsetup}(b)]. 
Reservoir induced energy shifts have already been observed experimentally and have been exploited to create on-demand potentials for polaritons via a spatially varying laser pump~\cite{Wertz2010,Tosi2012,Askitopoulos2013,Cristofolini2013,Gao2015,Pieczarka2021,estrecho2019,pieczarka2022}. 
Our theory of polaronic polaritons goes further and suggests that the very character of the polariton quasiparticle can be controlled by varying the reservoir density. 

To describe the many-body polariton-reservoir system, we formulate a novel Green’s function approach involving one and two-body correlators that allows us to incorporate the exact low-energy scattering between polaritons and reservoir excitons for arbitrary spin. 
In particular, it captures the coherent scattering of excitons out of the ``bright’’ (light-coupled) state, leading to a polaron quasiparticle that is a superposition of photon, bright exciton and scattered dark excitons (dressing cloud). 
This provides a natural mechanism for reducing the light-matter coupling with increasing exciton density. Moreover, unlike  conventional theories~\cite{schmittrink1985,rochat2000,glazov2009}, it does not invoke any Pauli blocking between the electrons and holes within the excitons. Thus, we expect it to dominate in typical experiments which are conducted below the Mott density.

Figure~\ref{fig:overallsetup}(b) summarizes the key implications of our polaronic polariton quasiparticles for the optical spectrum. We find that the upper and lower polaritons (UP and LP) are affected differently by the interactions with the medium, such that the UP polaron generally broadens with increasing density while the LP polaron shifts upwards in energy. This results in an effective reduction in the Rabi splitting between polariton branches, which mirrors the diminished light-matter coupling in the polarons and which is consistent with observations in experiment~\cite{Houdre1995,Rhee1996}. Crucially, this  differs from the standard picture of saturation since the loss of the Rabi splitting is primarily accomplished by the disappearance of the UP quasiparticle rather than the merger of two quasiparticle branches. We show that this result can explain recent experimental measurements of polariton interactions in transition metal dichalcogenides (TMDs)~\cite{stepanov2021}.

We furthermore see the appearance of the biexciton polariton (BP) at lower energies, which exhibits a pronounced Zeeman splitting since it relies on the presence of the biexciton bound state and is thus sensitive to the spin composition of both the reservoir and the polariton. This is associated with resonantly enhanced opposite-spin interactions that emerge from the dark-exciton-analog of a polariton Feshbach resonance~\cite{Wouters2007,Vladimirova2010,takemura2014,bleu2020}. Remarkably, we find that coherent BP polarons exist even though the excitonic medium is incoherent, and that such quasiparticles are stabilized by the strong coupling to light. 

The paper is organized as follows. In Section \ref{sec:model}, we outline the model and we present the Green's function formalism used in our work. We use this to solve the two-body problem exactly. 
In Section \ref{sec:results}, we generalize our result to the many-body case of an incoherent excitonic reservoir, and discuss the key properties of the resulting quasiparticle branches, the BP, LP, and UP polarons. We consider two particular experimental configurations in Section \ref{sec:expt}, both of which have recently been explored in microcavities containing a MoSe$_2$ monolayer as the active medium~\cite{stepanov2021, tan2022}. While those experiments were interpreted fully in terms of polaritons, we argue that an incoherent exciton reservoir could have played a key role in both cases.
We conclude in Section \ref{sec:conclusion}.

\section{Model and few-body properties}
\label{sec:model}

We consider a 2D semiconductor embedded in an optical microcavity, as depicted in Fig.~\ref{fig:overallsetup}(a). To describe an exciton polariton immersed in an incoherent dark excitonic medium, we employ the Hamiltonian $\hat{H} = \hat{H}_0  + \hat{V}$, which consists of the following single-particle and interaction parts (we set the system area and $\hbar$ to 1):
\begin{subequations}
\begin{align} \label{eq:H0}
\hat{H}_0 = & \sum_{\sigma} \bigg\{ \sum_{\k} \epsilon_{\k} \hat{x}^{\dagger}_{\k \sigma} \hat{x}_{\k \sigma} + \Delta_0 \hat{c}^{\dagger}_{\sigma} \hat{c}_{\sigma} \bigg\} \\
& + \Omega_0 \sum_{\sigma} \bigg[\hat{x}^{\dagger}_{0 \sigma} \hat{c}_{\sigma} + \hat{c}^{\dagger}_{\sigma} \hat{x}_{0 \sigma}\bigg] \nonumber 
\, ,\\ 
\hat{V} = & \sum_{\substack{\k, \k', \q, \\ \sigma, \sigma'}} \frac{v_{\sigma \sigma'}(\q)}{2} \hat{x}^{\dagger}_{\k \sigma} \hat{x}^{\dagger}_{\k' \sigma'}  \hat{x}_{\k'+\q \sigma'} \hat{x}_{\k-\q \sigma} \, .  \label{eq:Vint}
\end{align}
\end{subequations}
Here the bosonic operator \(\hat{x}^{\dagger}_{\k \sigma}\) creates an exciton with in-plane momentum \(\k\) and pseudospin \(\sigma = \up, \down\), corresponding to the spin of the electrons involved in the optical excitation of the exciton. 
We treat the excitons as featureless bosons, which is reasonable as long as the exciton binding energy is large compared to the energy scales of interest, as is typically the case in 2D semiconductors such as TMD monolayers and single GaAs quantum wells. 
We also assume that we are close enough to the band edge that the excitonic dispersion is given by \(\epsilon_{\k} = |\k|^2/2m_X \equiv k^2/2m_X\) with exciton mass $m_X$. All energies are with respect to the exciton energy at $\k=0$.

Equation~\eqref{eq:Vint} describes the interactions between excitions via the interaction potential  $v_{\sigma\sigma'}(\q)$. Importantly, this depends on the exciton spin, where we obviously require $v_{\up\down} = v_{\down\up}$ and we also assume that $v_{\up\up} = v_{\down\down}$. A complete description of the exciton-exciton interactions is presented in Sec.~\ref{subsec:int}.

For simplicity, we consider a single cavity mode of the microcavity, which is described by the photon creation operator \(\hat{c}^{\dagger}_{\sigma}\) with spin \(\sigma\) and detuning $\Delta_0$ with respect to the energy of the exciton at $\k =0$. The cavity photon is assumed to be at normal incidence to the semiconductor and thus it only couples to the exciton at $\k =0$ via the Rabi coupling \(\Omega_0\) in Eq.~\eqref{eq:H0}. The excitons with $\k \neq 0$ remain uncoupled to light and can thus naturally form a reservoir of optically dark excitons. Note that this scenario also applies to planar microcavities where the cavity photons have in-plane momenta, since the light-coupled excitons have $k$ much smaller than the typical momentum scales in the semiconductor. Hence, we can simply incorporate the effect of finite photon momentum by varying the light-matter detuning.

To connect our theory with the optical spectrum that is observed in experiments, we use the retarded photon Green's function which can be defined as
\begin{align} \label{eq:gctimedef}
    G_{C \sigma}(t) &= -i \theta(t) \braket{\hat{c}_{\sigma}(t) \hat{c}^{\dagger}_{\sigma}(0)} , 
\end{align}
where $\theta(t)$ is the Heaviside function and the average $\expval{\cdots}$ is taken over the state of the microcavity system assuming a small or negligible photon number, e.g., an incoherent medium of dark excitons.
Here we work within the Heisenberg picture such that we have time-dependent operators $\hat{c}_{\sigma}(t) = e^{i \hat{H} t} \hat{c}_{\sigma} e^{-i \hat{H} t}$. 
The Fourier transform, $G_{C \sigma}(\omega)$, of the photon Green's function then gives us access to the spectrum for each photon spin/polarization $\sigma$. In particular, the absorption is related to the spectral function, given by
\begin{equation}
\label{eq:specfuncdef}
    A_{\sigma}(\omega) = -\frac{1}{\pi} \Im G_{C \sigma}(\omega+i0), 
\end{equation}
while the transmission is proportional to $|G_{C \sigma}(\omega)|^2$~\cite{Cwik2016}. The frequency is shifted by an infinitesimal positive imaginary part due to the Heaviside function in Eq.~\eqref{eq:gctimedef}.

Importantly, while the Green's functions are diagonal in the spin basis, experiments are free to consider any polarization angle, which will in general probe the spectrum associated with both $G_{C \up}(\omega)$ and $G_{C \down}(\omega)$. For example, when the photon is of the form $\hat{c}_\theta=\cos(\theta)\hat{c}_\up+\sin(\theta)\hat{c}_\dn$, the spectral function becomes a weighted sum
\begin{equation}
\label{eq:angletrans}
    A_{\theta}(\omega) = \cos^2(\theta) A_{\uparrow}(\omega) + \sin^2(\theta) A_{\downarrow}(\omega),
\end{equation}
where $A_{\up}$ and $A_{\down}$ are the spin-resolved spectral functions defined in Eq.~\eqref{eq:specfuncdef}. 

\subsection{Non-interacting light-matter system}
\label{subsec:nonint}
In the absence of exciton-exciton interactions, the photon Green's function in Eq.~\eqref{eq:gctimedef} takes a particularly simple form. Due to the Rabi coupling, the Heisenberg equations of motion, $i \partial_t \hat{c}_{\sigma}(t) = [\hat{c}_{\sigma}(t), \hat{H}_0]$ and $i \partial_t \hat{x}_{0\sigma}(t) = [\hat{x}_{0\sigma}(t), \hat{H}_0]$, result in coupled equations for the photon and (zero-momentum) exciton operators. Fourier transforming to the frequency domain, we find the corresponding non-interacting Green's matrix (for details, see Appendix~\ref{app:eom})
\begin{align}
\mathbf{G}^{(0)}(\omega) & = 
\begin{pmatrix} 
\omega & -\Omega_0
\\ -\Omega_0 & \omega-\Delta_0 \end{pmatrix}^{-1}. \label{eq:G0matrix}
\end{align}
We have dropped the spin label since the Green's functions are spin independent in the absence of interactions. For ease of notation, in the following we define the non-interacting exciton Green's function at zero momentum,
\begin{align}
  \label{eq:GXnonint}
 G_X^{(0)}(\omega)\equiv\mathbf{G}^{(0)}_{11}(\omega)=\frac1{\omega-\frac{\Omega_0^2}{\omega-\Delta_0}},
\end{align}
as well as the corresponding photon Green's function
\begin{align}
  \label{eq:GCnonint}
  G_C^{(0)}(\omega)\equiv\mathbf{G}^{(0)}_{22}(\omega)=\frac1{\omega-\Delta_0-\frac{\Omega_0^2}{\omega}}.
\end{align}
We can also define the non-interacting exciton Green's function at finite momentum, $\k \neq 0$, which is uncoupled to light and thus has the form
\begin{align}
  \label{eq:GXnonint-k}
 G_X^{(0)}(\k,\omega)=\frac1{\omega-\epsilon_{\k}}.
\end{align}

Physically, the poles of the Green's matrix in Eq.~\eqref{eq:G0matrix} correspond to the eigenmodes of the system, i.e.,  the lower and upper polariton branches, with corresponding energies relative to that of the zero-momentum exciton  
\begin{equation}
\epsilon_{LP, UP}= \frac{1}{2} \big( \Delta_0 \mp \sqrt{\Delta_0^2 + 4\Omega_0^2} \big).
\end{equation}
It is often useful to consider the partial fraction decomposition of the non-interacting Green's function in terms of the eigenmodes. This yields
\begin{subequations}\label{eq:G0decompose}
\begin{align}
G^{(0)}_{C}(\omega) &=  \frac{|C_0|^2}{\omega - \epsilon_{LP}} +  \frac{|X_0|^2}{\omega - \epsilon_{UP}},\\
G^{(0)}_{X}(\omega) &=  \frac{|X_0|^2}{\omega - \epsilon_{LP}} +  \frac{|C_0|^2}{\omega - \epsilon_{UP}}.
\end{align}
\end{subequations}
Here we have defined the excitonic and photonic Hopfield coefficients, $X_0$ and $C_0$, via
\begin{equation}
    |X_0|^2 = \frac{1}{2}\bigg(1 + \frac{\Delta_0}{\sqrt{\Delta_0^2 + 4 \Omega_0^2}}\bigg),
\label{eq:hopfield}
\end{equation}
with $|C_0|^2=1-|X_0|^2$. This implies that the photonic fraction of the lower and upper polariton branches are $|C_0|^2$ and $|X_0|^2$, respectively. In the non-interacting case, the spectral function in Eq.~\eqref{eq:specfuncdef} then corresponds to Dirac delta function peaks at the corresponding energies, weighted by the photon fraction.

\subsection{Two-body problem}
\label{subsec:int}

We now consider the effect of exciton interactions in the two-body limit, which has the advantage that the corresponding modification of the Green's functions can be calculated exactly.
To proceed, we consider the scenario in which the medium consists of a single (dark) exciton of momentum $\Q$ and spin $\sigma'$. Hence, the averages in the Green's function are  taken with respect to the state $\ket{\Q\sigma'}=\hat{x}_{\Q\sigma'}^\dagger \ket{0}$.

\begin{figure}[tbp] 
\includegraphics[width=1\columnwidth]{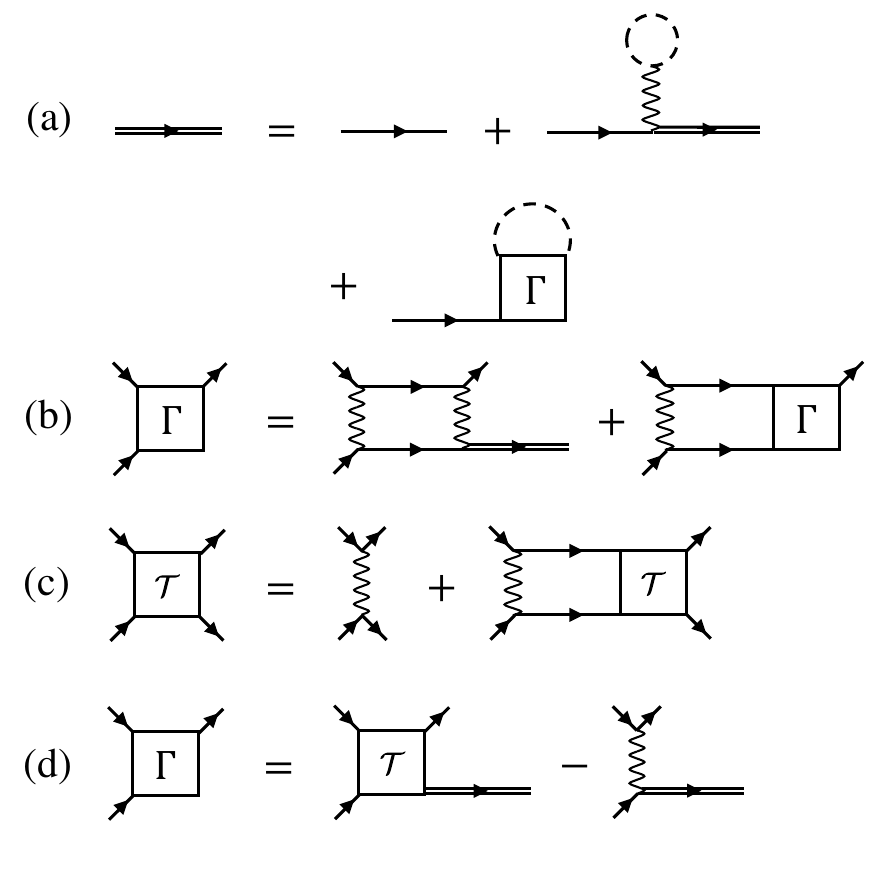}
\caption{Feynman diagrams for a spin-$\sigma$ polariton interacting with a single spin-$\bar\sigma$ reservoir exciton, where $\bar\sigma \neq \sigma$. The double-line propagators correspond to the interacting exciton Green's function, while the single horizontal lines represent its non-interacting counterpart (corresponding to Eqs.~\eqref{eq:GXnonint} and \eqref{eq:GXnonint-k} at zero and finite momentum, respectively). The exciton-exciton interaction potential is represented by a wiggly line, and the dotted loops indicate that the reservoir exciton has the same outgoing and incoming momenta. Panel (a) represents a combination of Eqs.~\eqref{eq:XGF2body} and \eqref{eq:step}, (b) corresponds to Eq.~\eqref{eq:gammap}, (c) is the standard definition of the two-body $T$ matrix (see Appendix~\ref{app:Tmat}), and (d) represents the relationship between the two-body correlator $\Gamma$ and the $T$ matrix in Eq.~\eqref{eq:Gamma}.}
\label{fig:diagrams}
\end{figure}

In analogy with the photon Green's function in Eq.~\eqref{eq:gctimedef}, the exciton Green's function reads
\begin{equation}
G_{X \sigma}(t) = -i \theta(t) e^{i \epsilon_\Q t}\braket{\hat{x}_{0 \sigma}e^{-i\hat{H}t} \hat{x}^{\dagger}_{0 \sigma}}, \label{eq:gxtimedef2body}
\end{equation}
where we have used $\hat{H}\ket{\Q\sigma'}=\epsilon_\Q\ket{\Q\sigma'}$. Note that we have explicitly included spin here, since the interaction part of the Hamiltonian is asymmetric with respect to spin. Upon Fourier transformation one obtains
\begin{equation}
G_{X \sigma}(\omega) = \braket{\hat{x}_{0\sigma} \frac{1}{\omega+\epsilon_\Q - \hat{H} + i0} \hat{x}^{\dagger}_{0 \sigma}}.
\label{eq:gxdef2body}
\end{equation}

Using the Born series expansion in terms of $\hat{V}$, one has the identity
\begin{align}\label{eq:usefulidentity}
\frac{1}{\omega+\epsilon_\Q-\hat{H}}&= \bigg(1+\frac{1}{\omega+\epsilon_\Q  - \hat{H}} \hat{V}\bigg) \frac{1}{\omega+\epsilon_\Q - \hat{H}_0},
\end{align}
which allows us to express the Green's function as
\begin{align}
G_{X\sigma}(\omega) =  G^{(0)}_{X}( \omega)
\left[1+\braket{\hat{x}_{0\sigma} \frac{1}{\omega+\epsilon_\Q - \hat{H}} \hat{V}  \hat{x}^{\dagger}_{0\sigma} }\right]\label{eq:XGF2body}. 
\end{align}
We see that the effect of interactions is encoded in the second term in brackets, which we now evaluate. The corresponding procedure is illustrated in Fig.~\ref{fig:diagrams} for the case of distinguishable excitons, the extension to indistinguishable excitons can be straightforwardly obtained.

\begin{widetext}
Using the definition of $\hat{V}$ in Eq.~\eqref{eq:Vint}, we have
\begin{align} 
  \braket{\hat{x}_{0\sigma} \frac{1}{\omega+\epsilon_\Q - \hat{H}} \hat{V} \hat{x}^{\dagger}_{0\sigma} }
    = \left[v_{\sigma\sigma'}(0)+\delta_{\sigma\sigma'}v_{\sigma\sigma}(\Q)\right]G_{X\sigma}(\omega)
    + \sideset{}{'}\sum_\k v_{\sigma\sigma'}(\k)\braket{\hat{x}_{0 \sigma} \frac{1}{\omega +\epsilon_\Q- \hat{H}} \hat{x}^{\dagger}_{\k \sigma} \hat{x}^{\dagger}_{\Q-\k \sigma'} \hat{x}_{\Q \sigma'} }, \label{eq:step}
\end{align}
where the prime on the sum implies that we have removed terms with $\k= 0$ and, for the case of $\sigma=\sigma'$, $\k=\Q$. Here, the first term on the right hand side corresponds to the Born approximation of scattering that is usually employed in the literature~\cite{tassone1999}, with the part proportional to $\delta_{\sigma\sigma'}$ due to the particle exchange that can occur when the medium exciton is identical to the light-coupled exciton.

To go beyond standard theories of polaritons, we need to include the second term on the right hand side of Eq.~\eqref{eq:step} which depends on a two-body correlator. Using Eq.~\eqref{eq:usefulidentity} we find that the corresponding sum satisfies the equation depicted in Fig.~\ref{fig:diagrams}(b):
\begin{align}
    \Gamma_{\sigma\sigma'}(\p,\Q,\omega)&\equiv\sideset{}{'}\sum_\k v_{\sigma\sigma'}({\k-\p})\braket{\hat{x}_{0 \sigma} \frac{1}{\omega +\epsilon_\Q- \hat{H}} \hat{x}^{\dagger}_{\k \sigma} \hat{x}^{\dagger}_{\Q-\k \sigma'} \hat{x}_{\Q \sigma'} }\nn \\
    &\hspace{-15mm}=G_{X\sigma}(\omega)\sideset{}{'}\sum_\k\frac{v_{\sigma\sigma'}({\k-\p})\left[v_{\sigma\sigma'}(\k)+\delta_{\sigma\sigma'}v_{\sigma\sigma}({\Q-\k})\right]}{\omega+\epsilon_\Q-\epsilon_{\k}-\epsilon_{\Q-\k}}
    +\sideset{}{'}\sum_\k v_{\sigma\sigma'}({\k-\p})\frac1{\omega+\epsilon_\Q-\epsilon_{\k}-\epsilon_{\Q-\k}}\Gamma_{\sigma\sigma'}(\k,\Q,\omega). 
    \label{eq:gammap}
    \end{align}
The formal solution of this equation can be obtained by considering its iteration. At $\p=0$, this takes the form
\begin{align}
    \Gamma_{\sigma\sigma'}(0,\Q,\omega)=G_{X\sigma}(\omega)\left[\mathcal{T}_{\sigma\sigma'}(\Q/2,\Q/2;\Q,\omega+\epsilon_\Q)-v_{\sigma\sigma'}(0)+\delta_{\sigma\sigma'}\left(\mathcal{T}_{\sigma\sigma}(\Q/2,-\Q/2;\Q,\omega+\epsilon_\Q)-v_{\sigma\sigma}(\Q)\right)\right].
    \label{eq:Gamma}
\end{align}
\end{widetext}
Here, the two-body $T$ matrix $\mathcal{T}(\p_1,\p_2;\Q,\omega)$, as illustrated in Fig.~\ref{fig:diagrams}(c) and discussed further in Appendix~\ref{app:Tmat}, is the sum of all scattering processes with center-of-mass momentum $\Q$, total energy $\omega$, and relative incoming (outgoing) momentum $\p_1$ ($\p_2$). Again, the last two terms on the right hand side originate from exchange of identical bosons. 

Comparing Eqs.~\eqref{eq:step} and \eqref{eq:Gamma}, we see that the inclusion of the two-body correlator in $\Gamma$ exactly replaces the contribution from the Born approximation by the full scattering $T$ matrix, and thus the exciton Green's function satisfies the Dyson equation~\cite{FetterBook}
\begin{equation}
    G_{X\sigma}(\omega) = G^{(0)}_{X}( \omega) + G^{(0)}_{X}( \omega) \Sigma_{\sigma}^{\text{2body}}(\omega) G_{X\sigma}(\omega),
\label{eq:gxdyson2body}
\end{equation}
with the self-energy
\begin{align}\label{eq:selfenergy2body}
\Sigma_{\sigma}^{\text{2body}}(\omega)=&\mathcal{T}_{\sigma\sigma'}(\Q/2,\Q/2;\Q,\omega+\epsilon_\Q)\nn \\ &+\delta_{\sigma\sigma'}\mathcal{T}_{\sigma\sigma}(\Q/2,-\Q/2;\Q,\omega+\epsilon_\Q) .
\end{align}
Equations \eqref{eq:gxdyson2body} and \eqref{eq:selfenergy2body} constitute an exact analytical result for the interacting Green's function in the two-body problem. Note that this is equivalent to the usual expression for the two-body Green's function from scattering theory~\cite{SakuraiQM} once we consider the fact that the self-energy is vanishingly small, scaling as inverse area.  
Hence $G_{X\sigma}$ on the right hand side of Eq.~\eqref{eq:gxdyson2body} can be replaced by $G^{(0)}_X$, such that we recover the standard relation $v G = \mathcal{T} G^{(0)}$. 
However, the advantage of our formulation is that it naturally connects to the Dyson equation in the many-body problem~\cite{FetterBook}, where the self-energy can instead scale with the density of the medium and thus be significant. Furthermore, our Green's function approach can be straightforwardly generalized to include higher body correlations such as those associated with triexcitons (three-exciton bound states)~\cite{levinsen2019,scarpelli2022probing}. 

In general, the $T$ matrix depends on the details of the interaction potential. However, a crucial simplification occurs for short-range potentials, such as the van der Waals type interaction between excitons, when the energy scale of interactions (set by the exciton binding energy) greatly exceeds all other relevant energy scales in the problem. In this case, the $T$ matrix is characterized by just a single parameter $\varepsilon_{\sigma \sigma'}$, and takes the form~\cite{bleu2020}
\begin{align} \label{eq:Tmat}
\mathcal{T}_{\sigma\sigma'}(\p_1,\p_2;\Q,\omega)&\simeq T_{\sigma \sigma'}(\Q, \omega )
\equiv \frac{4 \pi}{m_X \ln (-\frac{\varepsilon_{\sigma \sigma'}}{\omega - \epsilon_{\Q}/2})},
\end{align}
where $T_{\sigma \sigma'}(\Q, \omega )$ is the universal low-energy scattering $T$ matrix in 2D~\cite{adhikari1986quantum}.
To logarithmic accuracy, the energy scale $\varepsilon_{\sigma \sigma'}$ can be associated with a characteristic scale of the interactions. For \(\sigma = \sigma'\), the energy \(\varepsilon_{\up\up} = \varepsilon_{\down\down} \equiv \eb \) is  of order the exciton binding energy and describes the low-energy scattering between same-spin excitons. 
On the other hand, for \(\sigma \neq \sigma'\), the low-energy scattering is controlled by the existence of an $\up\down$ bound state---the biexciton---and thus the energy $\varepsilon_{\up\down}$ corresponds to the biexciton binding energy \(\varepsilon_{B}\). We emphasize that the validity of the analytical formula \eqref{eq:Tmat} for $\sigma=\sigma'$ is supported by microscopic calculations which include the composite nature of the excitons \cite{Schindler2008,Li2021}, while similar analytic formulas for polariton-electron scattering have also been verified by microscopic calculations~\cite{Li2021PRB,Li2021PRL,Kumar2023}.

The energy dependence of the $T$ matrix is a key element of our low-energy theory that is absent in descriptions of polariton and exciton interactions 
based on the Born approximation~\cite{carusotto2013review,tassone1999}. In particular, it implies that the relative strengths of the polariton-polariton, polariton-exciton and exciton-exciton interactions are not only dependent on the exciton fraction but are also affected by the collision energies appearing in the $T$ matrix. For bare 2D excitons, the two-body $T$ matrix vanishes logarithmically in the limit of zero momenta because the collision energy vanishes \cite{adhikari1986quantum}.
By contrast, the strength of polariton-polariton and polariton-exciton scattering remains sizeable at low momenta due to the shifted collision energy in the presence of strong light-matter coupling~\cite{bleu2020}. Finally, our low-energy approach can capture the broadening in the spectrum due to matter interactions since the structure of the $T$ matrix can lead to an imaginary part in the self-energy.

\section{Polaron quasiparticles}
\label{sec:results}

We now turn to the many-body problem and discuss how the interaction with a dark excitonic medium can strongly influence the behavior of exciton polaritons due to the formation of polaron quasiparticles. Our predictions include non-trivial shifts of the lower and upper polariton energies, modifications of their exciton fraction and associated light-matter coupling strength, saturation of their Rabi splitting, and even the emergence of new light-matter-coupled quasiparticles in the spectrum.

\begin{figure}[tbp] 
\includegraphics[width=1.02\columnwidth]{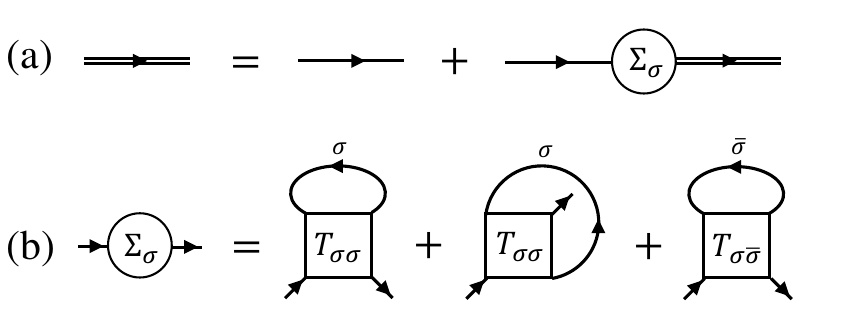}
\caption{Feynman diagrams for the many-body problem of a spin-$\sigma$ polariton interacting with a dark excitonic medium.
Panel (a) represents Dyson's equation~\eqref{eq:gxdyson} for the interacting exciton Green's function in terms of non-interacting Green's functions, where the effect of interactions is encoded in the spin-dependent self-energy $\Sigma_{\sigma}$. Panel (b) represents Eq.~\eqref{eq:selfenergytmat} and shows how the self-energy involves interactions with both same-spin and opposite-spin ($\bar{\sigma} \neq \sigma$) reservoir excitons. These are described by the low-energy $T$ matrices $T_{\sigma \sigma}$ and $T_{\sigma \bar{\sigma}}$, respectively, where there is an extra same-spin diagram due to particle exchange.
}
\label{fig:diagramsmb}
\end{figure}

Similarly to the exact two-body formulation in Eq.~\eqref{eq:gxdyson2body}, the effects of the dark medium  are all encoded in the self-energy $\Sigma_\sigma(\omega)$, such that the exciton Green's function satisfies the Dyson equation [Fig.~\ref{fig:diagramsmb}(a)]~\cite{FetterBook}
\begin{subequations}
\begin{align}\label{eq:gxdyson}
  G_{X\sigma}(\omega) &= G^{(0)}_{X}(\omega) + G^{(0)}_{X}(\omega) \Sigma_{\sigma}(\omega) G_{X\sigma}(\omega) \\
  & =  \frac{1}{\omega - \Sigma_{\sigma}(\omega) - \frac{\Omega_0^2}{\omega - \Delta_0}}.
\end{align}
\end{subequations}
This modifies the Green's matrix in Eq.~\eqref{eq:G0matrix} to the spin-dependent interacting form 
\begin{align}
\label{eq:greenmatrix}
\mathbf{G}_\sigma(\omega) & =  
\begin{pmatrix} 
\omega-\Sigma_\sigma(\omega) & -\Omega_0
\\ -\Omega_0 & \omega-\Delta_0 \end{pmatrix}^{-1} ,
\end{align}
where the interaction with the medium only affects the excitonic component. Thus we obtain the spin-dependent photon Green's function 
\begin{align} \label{eq:gc}
  G_{C \sigma}(\omega) 
  &= \frac{1}{\omega - \Delta_0 - \frac{\Omega_0^2}{\omega - \Sigma_{\sigma}(\omega)}}.
\end{align}

Before proceeding to consider the specific form of the self-energy that arises from interactions with a dark excitonic reservoir, let us first make some general observations that are valid for any medium, dark or otherwise. Provided the imaginary part of the self-energy is sufficiently small such that the upper and lower polaritons remain well-defined quasiparticles, we can treat the effect of the medium by considering their modified quasiparticle properties. The new medium-dressed and light-matter-coupled quasiparticles may be found by considering the poles of the Green's functions, i.e.,  we require $\Re[G^{-1}_{C,X\sigma}(\omega_p)]=0$~\footnote{Note that, in general, the poles of the two Green's functions can be slightly different due to the imaginary part of the self-energy. However, the difference is typically very small, for instance close to zero detuning it is $\simeq \frac12\Gamma_\sigma^2/\Omega_\sigma$.}. For frequencies close to such a pole, we then have
\begin{equation}
\label{eq:gxmod}
G_{X \sigma}(\omega) \simeq \frac{Z_{\sigma}}{\omega - \Delta_{X\sigma} - \frac{\Omega_\sigma^2}{\omega - \Delta_0} + i \Gamma_{\sigma}},
\end{equation}
and
\begin{equation}
\label{eq:gcmod}
G_{C \sigma}(\omega) \simeq \frac{1}{\omega - \Delta_0 - \frac{\Omega_\sigma^2}{\omega - \Delta_{X\sigma} + i \Gamma_{\sigma}}}.
\end{equation}
Here we have introduced the quasiparticle residue $Z_{\sigma}$, a key quasiparticle parameter that is defined as~\cite{FetterBook}
\begin{align} \label{eq:residue}
  Z_{\sigma} &= \bigg(1 - \bigg[\frac{\partial \Re[\Sigma_{\sigma}(\omega)]}{\partial \omega}\bigg]_{\omega = \omega_p}\bigg)^{-1},
\end{align}
and satisfies $0 \leq Z_\sigma \leq 1$. 
Comparing Eqs.~\eqref{eq:gxmod} and \eqref{eq:gcmod} with the non-interacting Green's functions in Eqs.~\eqref{eq:GXnonint} and \eqref{eq:GCnonint}, we find that $Z_\sigma$ changes the weight of the exciton Green's function and leads to the effectively reduced light-matter coupling
\begin{align}\label{eq:renormRabi}
  \Omega_\sigma=\sqrt{Z_\sigma} \Omega_0.
\end{align}

Physically, $Z_\sigma$ corresponds to the fraction of the matter component of the polaron quasiparticle that remains in the bright $\k=0$ exciton, while the remaining fraction ($1-Z_\sigma$) is in the polaronic dressing cloud composed of dark $\k \neq 0$ excitons and scattered medium particles. 
This physics relies on the frequency dependence of the self-energy, since a constant self-energy simply gives $Z_\sigma=1$ according to Eq.~\eqref{eq:residue}.
The interactions with the medium also cause the exciton resonance to shift and broaden according to
\begin{align}\label{eq:DeltaXsigma}
        \Delta_{X\sigma} &= \omega_p+Z_{\sigma}\left(\Re[\Sigma_{\sigma}(\omega_p)]-\omega_p\right), \\ \label{eq:Gammasigma}
       \Gamma_{\sigma} &= - Z_{\sigma} \Im[\Sigma_{\sigma}(\omega_p)],
\end{align}
respectively. 
In particular, we see that whether the exciton strongly shifts or broadens depends sensitively on whether the self-energy is predominantly real or imaginary. The former typically happens when the quasiparticle is detuned from a continuum of states, which is the case for the lower polariton. Conversely, the latter occurs when the quasiparticle sits in a continuum of states, as is typically the case for the upper polariton due to its upwards shift from the bare exciton line. Hence, the medium typically shifts the lower polariton while broadening the upper polariton, leading to a modified Rabi splitting, as illustrated in Fig.~\ref{fig:overallsetup}(b). For most parameter regimes, the Rabi splitting is reduced by the interaction with the medium; however, we show below that it is even possible to enhance the splitting between lower and upper polaritons in certain cases, even though the underlying light-matter coupling is reduced as per Eq.~\eqref{eq:renormRabi}. 

Let us contrast our ideas with the conventional treatment of polaritons in a (dark) excitonic reservoir~\cite{ferrier2011, Walker2017,estrecho2019,Fontaine2021}. Here it is assumed that the shifts of the lower and upper polaritons have the simple mean-field form
\begin{align}\label{eq:meanfield}
  \Delta_{LP}=g|X_0|^2 n, \qquad
  \Delta_{UP}=g|C_0|^2 n,
\end{align}
where $n$ is the reservoir density and $g$ is a constant exciton-exciton interaction strength, such as might be obtained in the Born approximation. This corresponds to taking the constant self-energy $\Sigma_\sigma(\omega) = g n$ and assuming $g n/\Omega_0\ll1$. Expressions like in Eq.~\eqref{eq:meanfield} are often used when modeling effective reservoir-induced potentials for an exciton-polariton superfluid.
However, this approach does not properly describe the full energy-dependent scattering of excitons and thus misses the polaronic physics we describe here. In particular, it gives $Z_\sigma=1$ and it does not capture the different interactions for the upper and lower polaritons. For instance, close to zero detuning where $|X_0|^2 = |C_0|^2 = 1/2$, Eq.~\eqref{eq:meanfield} predicts that both branches shift equally and thus the Rabi splitting remains constant. Therefore, in order to explain any reduction of the Rabi splitting between upper and lower polariton branches (i.e., saturation), one must invoke ``phase-space filling'' due to the fermionic nature of the electronic constituents of the excitons~\cite{schmittrink1985}. By contrast, our polaron theory naturally yields a saturation of the Rabi splitting, without the need to appeal to Pauli exclusion.

To make our discussion more concrete, we now consider a specific excitonic self-energy based on the exact solution of the two-body problem presented in Sec.~\ref{subsec:int} above. Specifically, we perform a weighted sum over the possible momenta and spins of the excitons in the reservoir, such that we obtain the self-energy depicted in Fig.~\ref{fig:diagramsmb}(b):
\begin{equation}
\label{eq:selfenergytmat}
    \Sigma_{\sigma}(\omega) = \sum_{\Q, \sigma'} (1+\delta_{\sigma \sigma'}) n_{\Q \sigma'} T_{\sigma \sigma'}(\Q, \omega + \epsilon_{\Q}),
\end{equation}
where $n_{\Q \sigma}$ is the occupation of a dark exciton with momentum $\Q$ and spin $\sigma$. Here, we have assumed that the relevant energy scales are such that we can take the universal expression for the $T$ matrix in Eq.~\eqref{eq:Tmat}. This self-energy corresponds to a so-called ladder approximation~\cite{rath2013}. As such, it neglects higher order correlations between the impurity and multiple excitations of the dark excitonic medium, as is reasonable for an incoherent reservoir where $n_{\Q \sigma}$ is approximately a Boltzmann distribution.

A further important simplification arises from the strong light-matter coupling combined with the fact that the dark medium particles are much heavier than the polaritons themselves. This implies that the collision energy in Eq.~\eqref{eq:selfenergytmat} is naturally dominated by the Rabi coupling rather than the exciton kinetic energy. Neglecting $\epsilon_\Q$ in Eq.~\eqref{eq:selfenergytmat}, the self-energy takes the particularly simple analytic form
\begin{equation}
\label{eq:selfenergy}
    \Sigma_{\sigma}(\omega) \simeq \frac{2 \varepsilon_{n_{\sigma}}}{\ln{\left(\frac{\varepsilon_{X}}{-\omega}\right)}} + \frac{\varepsilon_{n_{\bar{\sigma}}}}{\ln{\left(\frac{\varepsilon_{B}}{-\omega}\right)}} ,
\end{equation}
where we have defined \(\varepsilon_{n_{\sigma}} = 4 \pi n_{\sigma}/m_{X}\), with $n_{\sigma}$ the density of dark excitons for a given spin $\sigma$. The factor 2 in the first term originates from the indistinguishability of same-spin excitons, while the second term involves opposite spin interactions with $\bar{\sigma}\neq \sigma$. Provided $\omega$ is not resonant with the biexciton, we find that Eq.~\eqref{eq:selfenergy} is an excellent approximation when $\sum_{\Q\sigma} n_{\Q \sigma}\Theta(\epsilon_\Q-\Omega_0)\ll n$, with $n$ the total density of the dark medium. 

\begin{figure*}[htbp] 
\includegraphics[width=2.15\columnwidth, keepaspectratio]{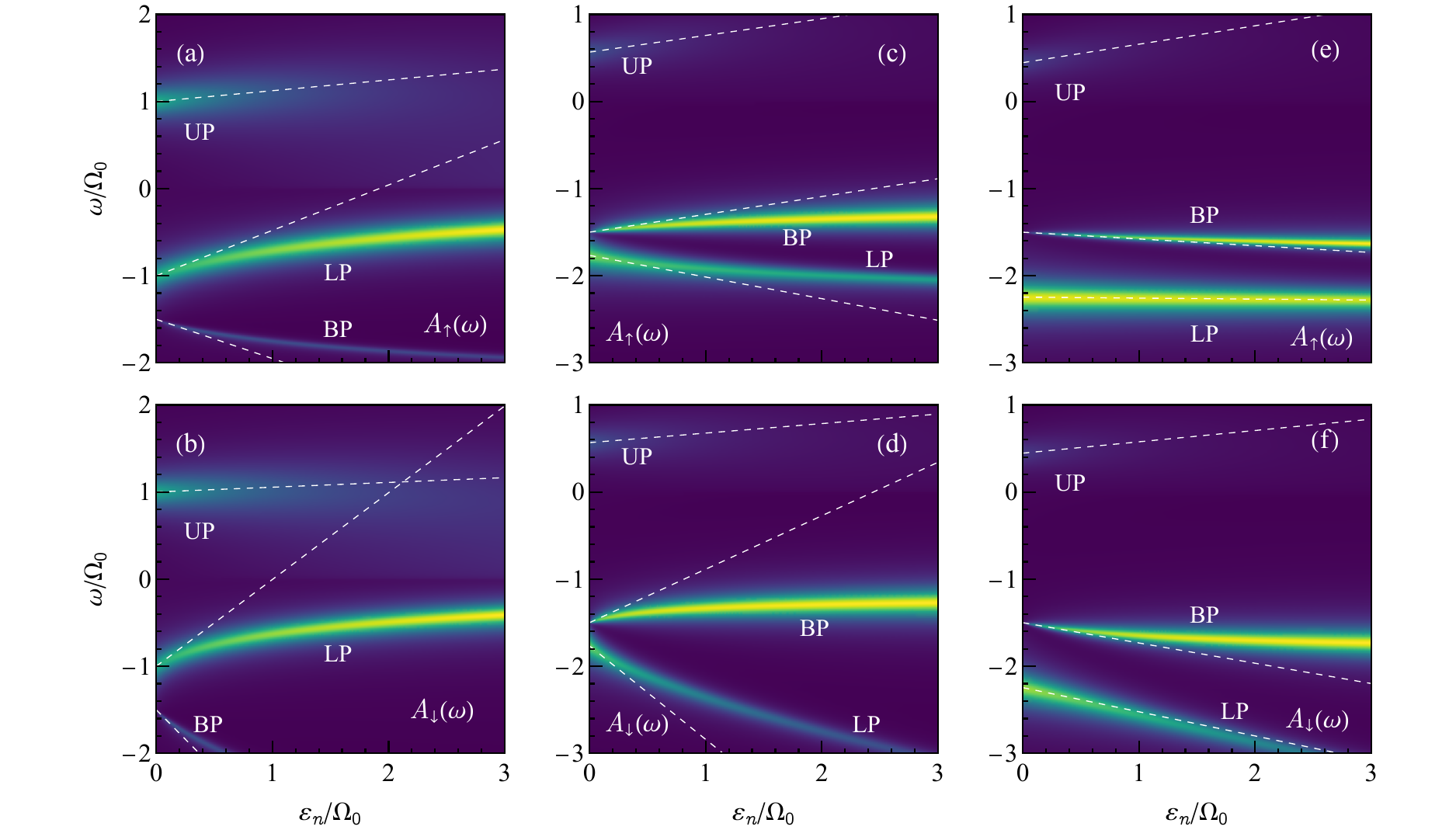}
\vspace{-10pt}
\caption{Spectral function as a function of total medium density, with $A_\sigma(\omega)$ evaluated using the self-energy in Eq.~\eqref{eq:selfenergy}. We consider a dark excitonic medium with $n_{\uparrow} = 3 n_\down$ and detunings (a, b) $\Delta_0 = 0$, (c, d) $\Delta_0 = -1.2 \Omega_0$, (e, f), $\Delta_0 = -1.8 \Omega_0$. The dashed white lines represent the corresponding low-density approximations for the energies of the UP, LP and BP polaron branches (Eqs.~\eqref{eq:UPupdown}, \eqref{eq:LPupdown}, and \eqref{eq:biexcFeature}, respectively). We have used parameters relevant to an MoSe$_2$ monolayer, where $\Omega_0\simeq 14$meV~\cite{stepanov2021}, such that \(\varepsilon_{B} = 1.5 \Omega_0\) and \(\varepsilon_{X}=34 \Omega_0\), with 
exciton and photon linewidths $\Gamma_X = \Gamma_C = 0.15 \Omega_0$ \cite{footnotelinewidth}.}
\label{fig:enplots}
\end{figure*} 

Using our analytic expression for the self-energy, Eq.~\eqref{eq:selfenergy}, we can immediately plot the evolution of the $\up$ and $\down$ spectral functions with increasing total reservoir density, as parameterized by $\varepsilon_n \equiv 4 \pi n/m_X$. The result is shown in Fig.~\ref{fig:enplots} for a reservoir with spin proportions $n_\up = 0.75 n$ and $n_\down = 0.25 n$. We see that each spectrum features three lines, rather than two, and that the two lower branches can either blueshift or redshift with increasing density. Furthermore, the corresponding lines in $A_\up$ and $A_\down$ can be Zeeman split with respect to each other. We now discuss each of these features in detail.

\subsection{Lower polariton polaron}
\label{subsec:lp}

We first consider the polaron quasiparticle that continuously evolves from the lower polariton with increasing medium density. This lower polariton polaron can be characterized by its medium-induced energy shift, as well as the relative weights of its excitonic and photonic components. 
In the following, we assume that we are away from any biexciton resonance, which will be discussed in Section \ref{subsec:bxp} when we consider the biexciton polariton. 

\subsubsection{Energy shifts}

To understand the behavior of the LP polaron energy, we start by focusing on the low-density limit, where we can obtain analytic expressions. In this regime, the polariton is only weakly perturbed by its interaction with the medium, and consequently we can evaluate the self-energy in Eq.~\eqref{eq:selfenergy} at $\epsilon_{LP}$. 
We then obtain the in-medium LP polaron energy by a series expansion around the unperturbed value, yielding (see Appendix \ref{app:applowdens} for details)
\begin{align}
\epsilon_{LP\sigma}(n)
& \simeq \epsilon_{LP}+|X_0|^2\left[\frac{2 \varepsilon_{n_{\sigma}}}{\ln \left( \frac{\varepsilon_{X}}{|\epsilon_{LP}|} \right)} + \frac{\varepsilon_{n_{\bar{\sigma}}}}{\ln{ \left( \frac{\varepsilon_{B}}{|\epsilon_{LP}|} \right)} }\right].
\label{eq:LPupdown}
\end{align}
This expression is plotted in Fig.~\ref{fig:enplots} and is seen to match the numerical results well at low density. It is linear in the density of the reservoir and proportional to the excitonic fraction. 
We therefore see that the low-density limit of our theory superficially resembles the commonly used mean-field expression of Eq.~\eqref{eq:meanfield}. However, in Eq.~\eqref{eq:LPupdown}, the interaction strength depends on the light-matter coupling via $\epsilon_{LP}$, which differs from the constant $g$ in the standard treatment. 
As we now discuss, this qualitative difference can have profound effects. 

According to Eq.~\eqref{eq:LPupdown}, the LP energy shift can involve dark excitons of the same and  opposite spin. While the former always leads to a blueshift, since the exciton binding energy exceeds any other relevant scale in the problem, the latter only leads to a blueshift if the lower polariton lies above the biexciton, $\epsilon_{LP}>-\varepsilon_B$ [Fig.~\ref{fig:enplots}(a,b)]; otherwise it leads to a redshift. Therefore, depending on the spin proportions of the reservoir and on the precise detuning, it is possible for the lower polariton to have an overall redshift with increasing density, as illustrated in Fig.~\ref{fig:enplots}(c-f). This prediction is qualitatively different from that of the conventional picture, which assumes that interactions are determined by lowest order exchange processes and phase-space filling \cite{renucci2005, leyder2007, glazov2009}, and only ever predicts blueshifts of the lower polariton. While the redshift leads to an increase of the splitting between upper and lower polaritons, we caution that this should not be interpreted as an increased light-matter coupling, since in this configuration a third quasiparticle peak, the biexciton polariton, appears in the spectrum above the lower polariton (see Fig.~\ref{fig:enplots} and Section \ref{subsec:bxp}). 

Furthermore, in the regime $|\epsilon_{LP}| \sim \varepsilon_{B}$, as is often the case in 2D TMDs or single-quantum-well semiconductors, the energy shift (blue or red) due to opposite-spin interactions can strongly dominate. This effectively corresponds to a dark-exciton analog of the polariton Feshbach resonance~\cite{Wouters2007,Carusotto2010,takemura2014,bleu2020}, where the polariton-exciton interaction is resonantly enhanced due to the biexciton. Notably, a strong polarization dependence of the lower polariton peak was recently observed in a 2D TMD~\cite{stepanov2021}, with the strongest blueshift occurring for linear polarization in accordance with this prediction. This behavior cannot be explained by the usual theory of saturation (based on the fermionic constituents of the excitons)~\cite{stepanov2021}, but it is well captured by our polaron theory --- see Section \ref{subsec:ComparisonStepanov}.

By comparing the $A_\up$ and $A_\down$ spectra in Fig.~\ref{fig:enplots}, we also see that when the reservoir is spin imbalanced, the $\up$ and $\dn$ lower polariton polarons are split. This splitting can be interpreted as an effective Zeeman splitting, induced by interactions with the exciton medium. At low density, the size of the splitting is
\begin{align}\label{eq:LPzeeman}
\epsilon_{LP\up}-\epsilon_{LP\dn}\simeq \frac{4\pi|X_0|^2 }{m_X} 
\frac{\ln\left(\frac{\varepsilon_{B}^2}{|\epsilon_{LP}|\varepsilon_{X}}\right)}{\ln(\frac{\varepsilon_X}{|\epsilon_{LP}|})\ln(\frac{\varepsilon_{B}}{|\epsilon_{LP}|})}
\left(n_\up-n_\dn\right),
\end{align}
where the sign and strength of the prefactor in Eq.~\eqref{eq:LPzeeman} are dependent on $\Delta_0$ and $\Omega_0$ through $\epsilon_{LP}$. 
Such a reservoir-induced Zeeman splitting of the lower polariton has recently been observed in a GaAs-based pillar microcavity~\cite{Real2021}, and a similar interaction-induced Zeeman splitting for excitons has been reported in WSe$_2$/MoSe$_2$ bilayers \cite{Srivastava2021}.

Beyond the low-density limit, the LP polaron energy is obtained from the poles of the photon Green's function, Eq.~\eqref{eq:gc}, which must be evaluated numerically. In the case where $\epsilon_{LP} > -\varepsilon_{B}$, there exist two limiting cases for the high-density behavior of the lower polariton branch: If $\Delta_0 \leq 0$, $\epsilon_{LP\sigma}(n)$ asymptotically approaches the photon energy, as shown in Fig.~\ref{fig:enplots}(a,b), while for $\Delta_0 > 0$ it is instead bounded from above by the exciton energy, as can be seen from the form of the self-energy in Eq.~\eqref{eq:selfenergy}. Conversely, if $\epsilon_{LP} < -\varepsilon_{B}$ as in Fig.~\ref{fig:enplots}(c-f), the LP polaron eventually becomes completely matter dominated, and its energy satisfies $\epsilon_{LP\sigma}(n)=\Sigma_\sigma(\epsilon_{LP\sigma}(n))$ in the high-density limit.

\subsubsection{Exciton and photon fractions}

Next, we discuss the impact of the reservoir on the matter and photonic components of the LP polaron. Similarly to the non-interacting case, Eq.~\eqref{eq:G0decompose}, one can show that Eqs.~\eqref{eq:gxmod} and \eqref{eq:gcmod} reduce to:
\begin{subequations}
\begin{align}
\label{eq:excweight}
G_{X \sigma} (\omega) &\simeq \frac{Z_{\sigma} |X_{\sigma}|^2}{\omega - \omega_{p}}, \\
G_{C \sigma} (\omega) &\simeq \frac{|C_{\sigma}|^2}{\omega - \omega_{p}},
\end{align}
\end{subequations}
in the vicinity of the LP polaron pole, where we have used the fact that the broadening $\Gamma_\sigma$ is negligible for the LP polaron. 
Here we have
\begin{equation}
\label{eq:LPpole}
    \omega_p= \frac{1}{2}\left(\Delta_0+\Delta_{X\sigma}-\sqrt{\Delta_{\sigma}^2+4\Omega^2_\sigma}\right),
\end{equation}
and the modified excitonic and photonic Hopfield coefficients, $X_{\sigma}$ and $C_{\sigma}$, given by 
\begin{equation}
\label{eq:xcrenorm}
    |X_{\sigma}|^2 = 1-|C_{\sigma}|^2=\frac{1}{2}\bigg(1 + \frac{\Delta_{\sigma}}{\sqrt{\Delta_{\sigma}^2 + 4 \Omega_{\sigma}^2}}\bigg),
\end{equation}
in terms of the modified detuning $\Delta_{\sigma} = \Delta_0 - \Delta_{X \sigma}$ and reduced light-matter coupling $\Omega_{\sigma}=\sqrt{Z_\sigma}\Omega_0$. These describe the weights of the LP polaron's excitonic and photonic components, where $|X_{\sigma}|^2$ corresponds to the \emph{total} exciton fraction involving both the bright $\k=0$ exciton and the scattered exciton in the polaronic dressing cloud.

We can gain further insight into the structure of the LP polaron by considering its behavior at low density. To leading order in $n$, the shifted exciton energy and (inverse) residue can be expressed as:
\begin{subequations} \label{eq:zdxapprox}
\begin{align} 
\Delta_{X\sigma}&\simeq  2\varepsilon_{n_\sigma} \frac{\ln \left(\frac{\varepsilon_X}{|\epsilon_{LP}|e}\right)}{\ln^2 \left(\frac{\varepsilon_X}{|\epsilon_{LP}|}\right)} +\varepsilon_{n_{\bar{\sigma}}} \frac{\ln \left(\frac{\varepsilon_B}{|\epsilon_{LP}|e}\right)}{\ln^2 \left(\frac{\varepsilon_B}{|\epsilon_{LP}|}\right)},
\\
Z_\sigma^{-1}&\simeq 1+ \frac{1}{|\epsilon_{LP}|} \left(\frac{2\varepsilon_{n_\sigma} }{\ln^2 \left( \frac{\varepsilon_X}{|\epsilon_{LP}|}\right)} + \frac{\varepsilon_{n_{\bar{\sigma}}}}{\ln^2 \left(\frac{\varepsilon_B}{| \epsilon_{LP}|}\right)} \right),
\end{align}
\end{subequations}
with $e$ Euler's number. In particular, like the LP polaron energy in Eq.~\eqref{eq:LPupdown}, we see that the bright exciton energy can either blueshift or redshift depending on the presence of opposite-spin dark excitons and the position of the biexciton. However, note that we only have the simple relation $\epsilon_{LP\sigma}(n) \simeq \epsilon_{LP} + |X_0|^2 \Delta_{X\sigma}$ when $|\epsilon_{LP}|$ is much smaller than the relevant interaction energy scales in the low-density regime. 
We also see that the residue $Z_\sigma$ is always smaller than unity, as expected, indicating the transfer of weight to the dressing cloud. 

\begin{figure}[tbp]
\includegraphics[width=1.05\columnwidth, keepaspectratio]{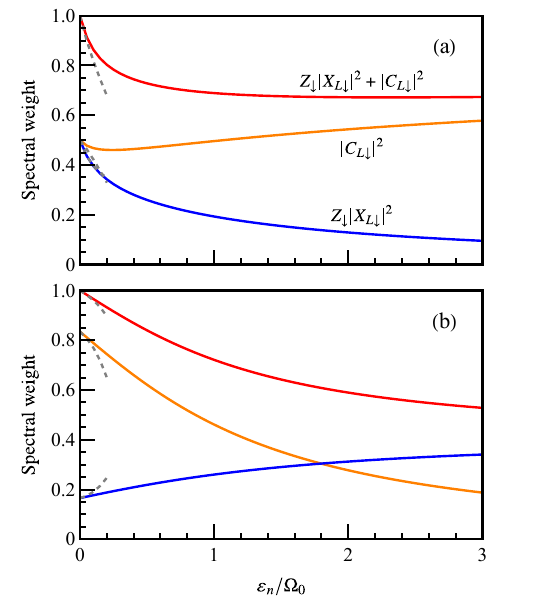}
\vspace{-10pt}
\caption{Total spectral weight (i.e., bare-polariton component) of the spin-$\down$ LP polaron (red) along with its excitonic (blue) and photonic (orange) fractions. Panel (a) corresponds to the parameters in Fig.~\ref{fig:enplots}(b) where $\Delta_0 = 0$ and the LP polaron blueshifts, while (b) corresponds to the parameters in Fig.~\ref{fig:enplots}(f) where $\Delta_0 = -1.8 \Omega_0$ and the LP polaron redshifts.
The dashed lines represent the low-density perturbative result obtained by combining Eqs.~\eqref{eq:zdxapprox} with \eqref{eq:xcrenorm}.}
\label{fig:residuesl}
\end{figure}

Figure \ref{fig:residuesl} displays our numerical results for the spectral weights of the LP polaron quasiparticle, corresponding to the ``bare'' polariton component, without the dressing cloud, as well as its exciton and photon fractions. We consider a reservoir with the same spin composition as in Fig.~\ref{fig:enplots} and we focus on the spin-$\down$ LP polaron where the energy shifts are most pronounced.
We see that the total spectral weight always decreases from 1 with increasing medium density owing to the decreasing $Z_\down$ and associated reduction in light-matter coupling. This is regardless of whether the lower polariton blueshifts (a) or redshifts (b). Furthermore, the low-density behavior is well captured using Eq.~\eqref{eq:zdxapprox} in Eq.~\eqref{eq:xcrenorm}.

The evolution of the relative photon and exciton fractions, $|C_{\down}|^2$ and $Z_\down |X_{\down}|^2$, depends on the interplay between $Z_\down$ and the shift in the exciton energy $\Delta_{X\down}$. In Fig.~\ref{fig:residuesl}(a), we see that both fractions initially decrease at low reservoir density, thus being dominated by the loss of weight to the dressing cloud. But eventually the photon fraction $|C_{\down}|^2$ increases with increasing density since the exciton energy blueshifts, resulting in a more negative effective detuning $\Delta_\down=\Delta_0-\Delta_{X\down}$. This effect can also be observed in Fig.~\ref{fig:enplots}(b) by noticing that the intensity of the LP polaron in the photon spectral function increases with increasing reservoir density. On the other hand, if $\Delta_{X\down}$ and the lower polariton redshifts, the photon fraction always decreases, as can be seen in Fig.~\ref{fig:residuesl}(b) and Fig.~\ref{fig:enplots}(f). 
In this case, the exciton fraction increases but only relatively slowly because of the strong loss of spectral weight due to enhanced exciton-exciton scattering near the biexciton resonance.

\subsection{Biexciton polariton}
\label{subsec:bxp}

In addition to the LP polaron, the spectral function $A_\sigma$ features another peak below the exciton that continuously evolves into the biexciton at vanishing medium density (see Fig.~\ref{fig:enplots}). The resulting polaron quasiparticle---the biexciton polariton---can be viewed as the neutral analog of the so-called trion polariton which can form when a polariton is immersed in an electron gas~\cite{sidler2017,tan2020,Emmanuele2020,Lyons2022}. Importantly, it only exists when the density of reservoir excitons with opposite spin $\bar\sigma$ is nonzero and thus its existence inherently relies on both light-matter coupling and on the reservoir. 
Signatures of such a quasiparticle have already been observed in pump-probe measurements on semiconductor quantum wells~\cite{Saba2000,Neukirch2000} and indicate physics beyond the usually employed mean-field theories. 

In this section, we provide insight into the biexciton polariton by presenting analytical results for the BP polaron energy in the low-density limit, using the approximate self-energy Eq.~\eqref{eq:selfenergy}. 
We then show that the momentum distribution of medium excitons, captured in Eq.~\eqref{eq:selfenergytmat}, can give rise to a substantial broadening of the spectral lines appearing below the bare biexciton energy, which has implications for the very existence of the biexciton polariton quasiparticle.

\subsubsection{Low-density limit}

We first assume that we have a low density of excitons and we consider frequencies where $\omega \sim-\varepsilon_B$. In this case, the term $\varepsilon_{n_{\bar{\sigma}}}/\ln{(-\varepsilon_{B}/\omega)}$ in the self-energy \eqref{eq:gc} is near divergent \footnote{There exists a spurious divergence in the self-energy if $\omega = -\varepsilon_{X}$, as a consequence of using the low-energy result beyond its limits of validity. It does not appear in the full four-body calculation~\cite{Li2021PRB}.}
such that we can ignore the contribution to the self-energy from excitons of spin $\sigma$. If the lower polariton is sufficiently far detuned from the biexciton energy, 
then we can straightforwardly expand the denominator of the photon Green's function in the vicinity of $\omega=-\varepsilon_{B}$, and we find an additional root corresponding to the biexciton polariton. In the low-density limit, the corresponding energy is (see Appendix \ref{app:applowdens})
\begin{align}\label{eq:biexcFeature}
     \epsilon_{BP \sigma}(n) \simeq -\varepsilon_{B}\left(1+  \frac{\varepsilon_{n_{\bar\sigma}}}{\varepsilon_{B} - \frac{\Omega_0^2}{\Delta_0 + \varepsilon_{B}}}
     \right).
\end{align}
Similarly to the LP polaron, we see that the energy is sensitive to the reservoir excitons' spin, and that there is an effective Zeeman splitting proportional to any spin imbalance $n_\up-n_\dn$ of the reservoir. Such a splitting can clearly be seen by comparing the spin-$\up$ and spin-$\down$ panels of Fig.~\ref{fig:enplots}.

Our perturbative result \eqref{eq:biexcFeature} also shows how the BP polaron energy can either redshift or blueshift depending on the relative position of the lower polariton. Specifically, we see that the biexciton polariton redshifts in the case where $\epsilon_{LP} > -\varepsilon_{B}$, as depicted in Fig.~\ref{fig:enplots}(a,b). In this regime, the LP and BP quasiparticles can be interpreted as repulsive and attractive polaron branches, respectively, where the former repels the surrounding medium particles, while the latter attracts them. This is a hallmark of the behavior of Bose and Fermi polarons in ultracold atomic gases~\cite{Scazza2022review}.

The situation where $\epsilon_{LP} < -\varepsilon_{B}$ is more subtle, since it is now possible for the photon to become resonant with the biexciton, $\Delta_0 = -\varepsilon_B$, in which case we see that the BP polaron energy is independent of density according to Eq.~\eqref{eq:biexcFeature}. This singular behavior implies that the BP polaron either redshifts or blueshifts depending on whether $\Delta_0 <-\varepsilon_{B}$ or $\Delta_0 >-\varepsilon_{B}$, respectively, as shown in Fig.~\ref{fig:enplots}.   

In the resonant case where $\epsilon_{LP}=-\varepsilon_{B}$, the lower polariton and biexciton are naturally entwined, which modifies the pole expansion of the photon Green's function in the vicinity of $\omega=-\varepsilon_B$. This gives rise to two branches with energies (see Appendix \ref{app:applowdens})
\begin{align}\label{eq:resonance}
    \epsilon_{\pm, \sigma}(n) &\simeq 
    -\varepsilon_B\left(1 \pm \sqrt{\frac{\varepsilon_{n_{\bar\sigma}}}{\varepsilon_B}\frac{1+\Delta_0/\varepsilon_B}{2+\Delta_0/ \varepsilon_{B}}}\right),
\end{align}
which scale as $\sqrt{n}$ rather than being linear in density, in agreement with the behavior in Fig.~\ref{fig:qpexistence}(a). 
The same scaling was predicted in the situation where the medium was a coherent state of $\dn$ lower polaritons \cite{levinsen2019} instead of an excitonic reservoir. 
Note, further, that this scaling implies that the effective Zeeman splitting between these branches in this case is $\epsilon_{\pm \up}-\epsilon_{\pm \dn}\propto \sqrt{n_{\dn}}-\sqrt{n_\up}$. 

Our analytic expressions accurately describe the energy shifts in the low-density limit and even capture the qualitative behavior obtained from the full self-energy~\eqref{eq:selfenergytmat}. However, in order to obtain the linewidths, we need to go beyond the approximation of Eq.~\eqref{eq:selfenergy}. 

\subsubsection{Existence of the polaron quasiparticle}
\label{subsubsec:distribution}

\begin{figure}[tbp] 
\includegraphics[width=1.05\columnwidth, keepaspectratio]{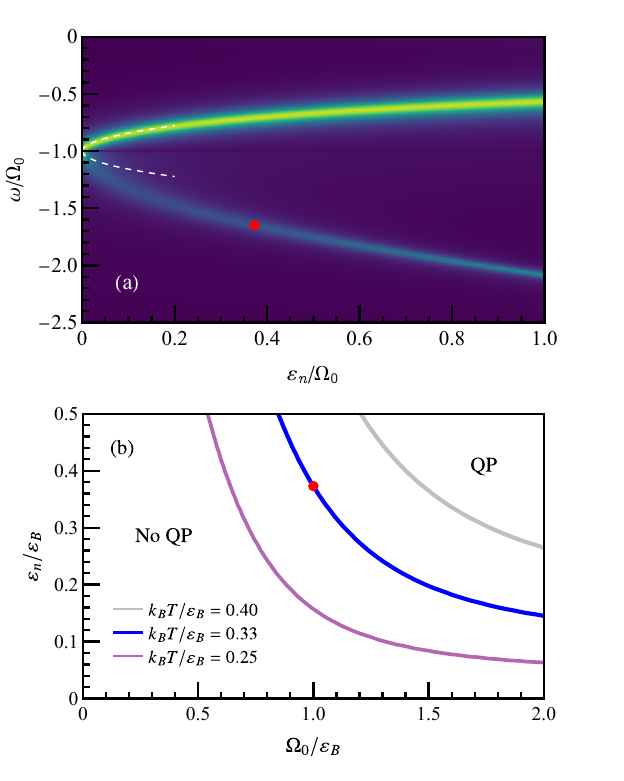}
\caption{(a): Spectral function for a fully polarized opposite-spin medium at effective temperature $k_{B} T/\varepsilon_{B} = 0.33 $ and $\Omega_0 = \varepsilon_B$ at resonance ($\epsilon_{LP} = - \varepsilon_{B}$). The white dashed lines are the corresponding low density energies given by Eq.~\eqref{eq:resonance}, and the red dot represents the point where the BP polaron becomes a well-defined quasiparticle. (b) Existence of the biexciton polariton quasiparticle at finite temperature, at resonance $(\epsilon_{LP} = -\varepsilon_{B})$. The lines represent the boundary where the quasiparticle start to be defined for different ratios of $k_{B} T/\varepsilon_{B}$. To calculate the boundaries, we used the self-energy given in Eq.~\eqref{eq:approxse}. Below the lines, there is no well-defined quasiparticle (no QP), while above the lines, there is a well-defined quasiparticle. The red dot marks the boundary for the ratio $\Omega_0/\varepsilon_B$ used in panel (a).}
\label{fig:qpexistence}
\end{figure}  
For an incoherent excitonic reservoir, the large spread of possible collision energies can substantially broaden the biexciton polariton, ultimately destroying the polaron quasiparticle. This effect arises 
from the fact that the biexciton resonance occurs when $\omega+\epsilon_\Q/2=-\varepsilon_B$ for a given momentum $\Q$ of a reservoir exciton, and thus there is not a single resonant frequency when there is a distribution of momenta in the reservoir. 
Indeed, this picture provides a natural explanation for why the biexciton resonance often features a substantial linewidth. 
The resulting broadening is not captured by the approximate self energy \eqref{eq:selfenergy} that we have considered so far, and hence we must instead turn to the self-energy in Eq.~\eqref{eq:selfenergytmat}, which explicitly takes the momentum distribution of the incoherent medium into account. 

To be concrete, in the following we focus on the case where the exciton reservoir and the polariton are of opposite spins, and we assume that the medium is sufficiently thermalized such that we can use a Boltzmann distribution with temperature $T$,
\begin{equation}
n_{\Q\bar\sigma} = \frac{\varepsilon_{n_{\bar\sigma}}}{2 k_{B} T} \exp (-\frac{\epsilon_{\Q}}{k_{B} T}).
\label{eq:boltzmann}
\end{equation}
With this distribution, we can now ask the question: for what combination of physical parameters does the biexciton polariton exist as a well-defined quasiparticle? To address this important point, we will use the following criterion: In order for a quasiparticle to exist, its energy must satisfy $\Re[G_{C \sigma}^{-1}(\omega_p)]=0$ \cite{FetterBook}. While we could equally well consider a similar criterion for the exciton Green's function, here we focus on the photon since the associated spectral function is closely related to experimental observables. 

For a Boltzmann gas, the self-energy~\eqref{eq:selfenergytmat} in the vicinity of $\omega\sim-\varepsilon_B$ is well approximated by the analytic expression \cite{mulkerin2022}
\begin{align}
\Sigma_{
\sigma}(\omega) \simeq & -2 \frac{\varepsilon_{n_{\bar{\sigma}}}}{k_{B} T} \varepsilon_{B} \exp (\frac{2 (\omega + \varepsilon_{B})}{k_{B} T}) \nonumber \\ 
&\times\bigg[\text{Ei}\bigg(-\frac{2 (\omega + \varepsilon_{B})}{k_{B} T}\bigg) + i \pi \theta(-\omega - \varepsilon_{B}))\bigg],
\label{eq:approxse}
\end{align}
where $\text{Ei}$ is the exponential integral.Thus, we see that the temperature-induced imaginary part of the self-energy is much stronger for energies immediately below the biexciton than above, as can be seen from the Heaviside function in Eq.~\eqref{eq:approxse}.
This asymmetry implies that the broadening induced by the finite temperature mostly affects the spectral function for $\omega<-\varepsilon_B$, as can clearly be seen in Fig.~\ref{fig:qpexistence}(a).

To investigate the existence of the polaron quasiparticle, we consider the resonant case, $\epsilon_{LP}=-\varepsilon_B$, where the effect of the biexciton is the most pronounced in the spectrum. In this case, according to Eq.~\eqref{eq:resonance}, there are two branches that emerge from the biexciton at low density, which both have mixed character of lower polariton and biexciton. However, the upper of these solutions lies above the biexciton and thus does not suffer from any broadening induced by finite-momentum excitons. Therefore, we consider the regimes of quasiparticle existence for the lower (attractive) branch, as illustrated in Fig.~\ref{fig:qpexistence}. 
We see that the polaron quasiparticle is destroyed by thermal fluctuations (i.e., the broad distribution in momenta) if the temperature scale $k_B T$ is sufficiently large compared to $\varepsilon_n$. Therefore, for a fixed temperature, the quasiparticle can be stabilized by increasing the reservoir density, as in Fig.~\ref{fig:qpexistence}(a). Furthermore, the coupling to the photon also favors the existence of a quasiparticle, and remarkably we see in Fig.~\ref{fig:qpexistence}(b) that a biexciton polariton quasiparticle can emerge even when no polaron quasiparticle exists in the absence of light-matter coupling. 

\subsection{Upper polariton polaron}
\label{subsec:up}

Finally, we consider the behavior of the upper polariton in the presence of a dark excitonic medium. As highlighted previously, the upper polariton sits in a continuum of exciton-exciton scattering states, causing the UP polaron to broaden more than shift. In contrast to the biexciton polariton above, this effect can be captured by the simplified self-energy in Eq.~\eqref{eq:selfenergy}, which becomes complex when $\omega > 0$. To clearly see this effect, we perform a perturbative expansion at low density  (for details, see Appendix~\ref{app:applowdens}), to find the medium-induced shift of the energy:
\begin{align}
    &\epsilon_{UP\sigma}(n) \nn \\
    &\, \simeq\epsilon_{UP}+|C_0|^2\left[\frac{2 \varepsilon_{n_{\sigma}}\ln \left(
    \frac{\varepsilon_{X}}{\epsilon_{UP}}\right)}{\ln^2\left( \frac{\varepsilon_{X}}{\epsilon_{UP}}\right)+\pi^2} + \frac{\varepsilon_{n_{\bar{\sigma}}}\ln\left(\frac{\varepsilon_{B}}{\epsilon_{UP}}\right)}{\ln^2\left(\frac{\varepsilon_{B}}{\epsilon_{UP}}\right)+\pi^2}\right].
    \label{eq:UPupdown}
\end{align}
Comparing this with the corresponding expression for the low-density shift of the LP polaron, Eq.~\eqref{eq:LPupdown}, we clearly see that the energy shift of the upper polariton is typically much smaller. For instance, at zero detuning, where $|X_0|^2=|C_0|^2$ and $|\epsilon_{LP}|=\epsilon_{UP}$, the difference is solely due to the presence of the factor $\pi^2$ of Eq.~\eqref{eq:UPupdown}, which for almost all experimentally relevant parameters will dominate the denominator. As a result, the Rabi splitting between the two branches is typically reduced, which mirrors the reduction in light-matter coupling, $\Omega_\sigma \simeq \sqrt{Z_\sigma}\Omega_0$.

\begin{figure}[tbp] 
\includegraphics[width=1.06\columnwidth, keepaspectratio]{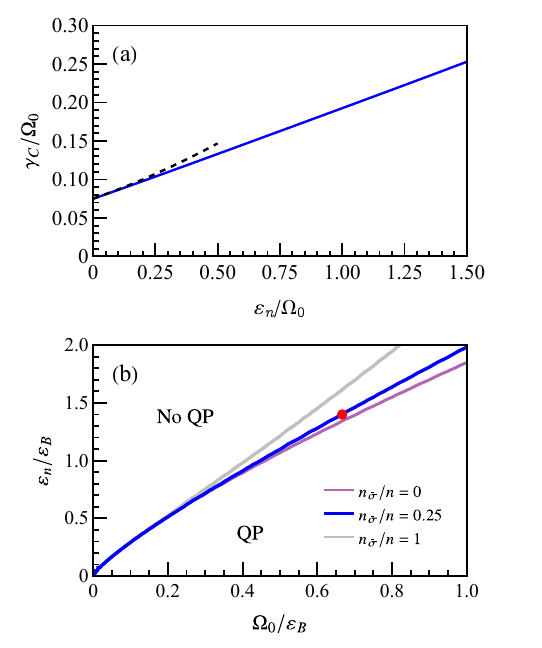}
\caption{(a) The half-width at half-maximum of the spin-$\down$ UP polaron (solid blue line) extracted from the spectral function in Fig.~\ref{fig:enplots}(a). At low-density, this matches $\gamma_C$ obtained from the pole expansion in Eq.~\eqref{eq:UPGCbroadenings} for the parameters in Fig.~\ref{fig:enplots}(a), where we have added $|X_{\sigma}|^2 \Gamma_C$ to account for the intrinsic linewidth (black dashed line).  
(b) Existence of the UP polaron quasiparticle for $\Delta_0 = 0$ and using Eq.~\eqref{eq:selfenergy}. The lines represent the boundary of where the quasiparticle (QP) is well defined, for different ratios $n_{\bar{\sigma}}/n$. The blue line corresponds to the ratio used in Fig.~\ref{fig:enplots}(a), where the loss of the quasiparticle occurs at $\varepsilon_n/\varepsilon_B \approx 1.4$ (red dot) for the particular $\Omega_0/\varepsilon_B$ considered.}
\label{fig:upexistence}
\end{figure} 

The associated exciton broadening \eqref{eq:Gammasigma} at low density takes the form
\begin{equation}
    \Gamma_{\sigma}\simeq
    \frac{2\pi \varepsilon_{n_{\sigma}}}{\ln^2\left( \frac{\varepsilon_{X}}{\epsilon_{UP}}\right)+\pi^2} + \frac{\pi\varepsilon_{n_{\bar{\sigma}}}}{\ln^2\left(\frac{\varepsilon_{B}}{\epsilon_{UP}}\right)+\pi^2}.
    \label{eq:Gammaupdown}
\end{equation}
For typical experimental parameters, this interaction-induced broadening is larger than the energy shift, and leads to a strongly broadened spectral function as clearly seen in Fig.~\ref{fig:enplots}.

Despite the significant broadening, we may still analyze the exciton and photon fractions of the UP polaron. Confining ourselves to the regime $\Gamma_{\sigma} 
\ll \Omega_0$, we perform an expansion around the quasiparticle pole which yields
\begin{subequations} \label{eq:UPgreen}
\begin{align}
G_{X \sigma} (\omega) &\simeq \frac{Z_{\sigma} |C_{\sigma}|^2}{\omega - \omega_{p} + i \gamma_X}, \\
G_{C \sigma} (\omega) &\simeq \frac{|X_{\sigma}|^2}{\omega - \omega_{p} + i \gamma_C},
\end{align}
\end{subequations}
where we have for the UP polaron
\begin{equation}\label{eq:UPpole}
    \omega_p\simeq \frac{1}{2}\left(\Delta_0+\Delta_{X\sigma}+\sqrt{\Delta_{\sigma}^2+4\Omega^2_\sigma}\right),
\end{equation}
and the definitions of $|X_{\sigma}|^2, |C_{\sigma}|^2$ are as in Eq.~\eqref{eq:xcrenorm} (but where $\Delta_\sigma$ and $\Omega_\sigma$ are now evaluated for the UP polaron). 
In the absence of intrinsic linewidths, $\Gamma_C$, $\Gamma_X$, the effective broadenings in Eq.~\eqref{eq:UPgreen}
are
\begin{subequations}
    \begin{align}
        \gamma_{X} &\simeq |C_{\sigma}|^2 \Gamma_{\sigma}, \label{eq:UPGXbroadenings}\\
        \gamma_{C} &\simeq|X_{\sigma}|^2 \frac{4\Omega_{\sigma}^2 \Gamma_{\sigma}}{  (\Delta_\sigma+\sqrt{\Delta_\sigma^2+4\Omega_\sigma^2})^2}. 
    \label{eq:UPGCbroadenings}
    \end{align}
\end{subequations}
Thus, we once again obtain exciton and photon Green's functions that resemble those in the non-interacting case,  Eq.~\eqref{eq:G0decompose}, with Hopfield coefficients appropriately reversed for the upper polariton.
Note that the approximate form of the photon Green's function requires $2\Gamma_{\sigma}/|\Delta_{\sigma} +\sqrt{\Delta_\sigma^2+4\Omega_\sigma^2}|\ll1$. This originates from the fact that when $\Gamma_\sigma$ increases, the pole of $G_{C\sigma}$ starts to differ from that of $G_{X\sigma}$. Interestingly, we see that when $\Delta_\sigma=0$ we have $\gamma_C\simeq\gamma_X$.

The resulting width of the UP polaron is shown in Fig.~\ref{fig:upexistence}(a), where we also include an intrinsic linewidth. We see that Eq.~\eqref{eq:UPGCbroadenings} captures the quasiparticle width at low densities, while it overestimates the broadening for larger densities. Furthermore, the width of the peak is relatively insensitive to the precise spin composition of the medium, since the logarithms in the self-energy, Eq.~\eqref{eq:selfenergytmat}, take similar values for positive frequencies.

The upper polariton can also cease to be a well-defined quasiparticle in a similar manner to the biexciton polariton. However, in the case of the UP polaron, this is predominantly due to interactions rather than temperature. This is illustrated in Fig.~\ref{fig:upexistence}(b), where we see that the quasiparticle existence depends strongly on $\varepsilon_n$ relative to the Rabi coupling, with larger densities leading to a less well-defined quasiparticle as the UP polaron is shifted further up into the exciton scattering continuum. On the other hand, similarly to the width of the UP polaron, the existence of the quasiparticle only weakly depends on the spin composition of the exciton medium.
The loss of the UP polaron quasiparticle is qualitatively different from the loss of strong coupling due to saturation of the emitters, since the lower and upper polariton polarons in our case do not merge with increasing density.

Finally, we stress that the broadening of the UP polaron considered here is qualitatively different from the broadening that generically happens when the upper polariton is shifted into the electron-hole continuum in a 2D semiconductor microcavity~\cite{Levinsen2019prr} (or, in the case of an organic material, when the upper polariton enters a continuum of molecular excitations~\cite{garciajomaso2023}). While the loss of the quasiparticle in our scenario is distinctly a many-body phenomenon, the other scenario depends solely on the properties of the fundamental excitations of the semiconductor. 

\section{Experimental scenarios}
\label{sec:expt}

\begin{figure*}[tbp] 
\vspace{-5pt}
\includegraphics[width=2.19\columnwidth, keepaspectratio]{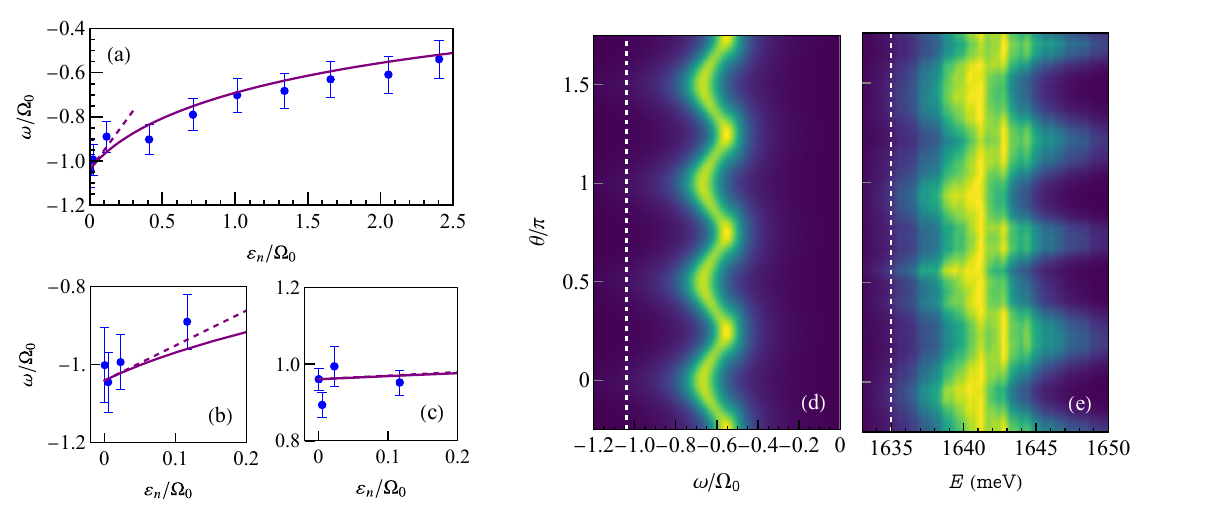}
\vspace{-10pt}
\caption{Comparison to experimental results in Ref.~\cite{stepanov2021} for (a-c) linear polarization and (d,e) varying polarization. (a) The observed blueshift of the lower polariton branch (blue dots) is compared to the peak of the spectral function calculated using Eq.~\eqref{eq:selfenergy} (line, solid) and to the low-density expression in Eq.~\eqref{eq:LPupdown} (purple line, dashed) \cite{stepanovfootnote}. (b,c) The low-density data compared to both Eq.~\eqref{eq:LPupdown} and Eq.~\eqref{eq:UPupdown} (dashed lines) and the theoretical predictions for the peaks (solid lines), for the lower and upper polariton branches, respectively. (d) Calculated spectral function at fixed density (corresponding to $\varepsilon_n = 2 \Omega_0$) as a function of polarization angle \cite{polarizeranglefootnote}. (e) Shows the corresponding experimentally observed transmission spectrum [as a result, the colour scales are not the same in panels (d) and (e)]. The white dotted line is the unshifted lower polariton energy (1635 meV corresponding to -1.04 $\omega/\Omega_0$), and the plot range is adjusted to be the same in both panels. Parameters: $|X_0|^2 = 0.48$, $\Omega_0 = 14.0 \pm 1.5 $ meV, $\varepsilon_{X} = 33.6 \Omega_0$ (470 meV \cite{dufferwiel2015}), $\varepsilon_{B} = 1.43 \Omega_0$ (20 meV \cite{hao2017}), exciton and photon linewidths $\Gamma_X = 0.1 \Omega_0$ and $\Gamma_C = 0.1 \Omega_0$ \cite{footnotelinewidth}.}
\label{fig:exptplot}
\end{figure*} 

We now consider two experimentally relevant scenarios of particular interest. In the first, we assume that the exciton reservoir is polarized along the same angle as the polariton and determine the spectral function as a function of polarization angle. This allows us to clearly identify the contributions to interactions due to same and opposite spin interactions, including the effect of the biexciton resonance, as well as the polarization-dependent reduction of the effective light-matter coupling. Second, we consider a $\down$ polariton in a fully polarized spin-$\up$ dark reservoir and we investigate signatures of the biexciton polariton as a function of the detuning. 

Such scenarios have recently been experimentally realized in MoSe$_2$ monolayers embedded in microcavities---see Refs.~\cite{stepanov2021} and \cite{tan2022}. In both cases, the medium was assumed to consist of bright polaritons, rather than dark excitons. However, we find that we can capture the main features of these experiments using our polaron theory with a dark excitonic reservoir, thus suggesting that dark excitons could have played a role. Therefore, in the following we will discuss these two experimental scenarios using parameters from the experiments.

\subsection{Polariton in a co-polarized exciton medium}
\label{subsec:ComparisonStepanov}

First, we apply our theory to the setup reported in Ref.~\cite{stepanov2021}, which investigated the response of the lower and upper polariton branches to changes in two parameters: the intensity of the pump laser, and the polarization of that laser. In order to establish a connection between the laser power and reservoir density, we assume that the dark exciton reservoir density is proportional to the laser pump power. We also assume that the reservoir is polarized in the same manner as the polariton~\footnote{We note that the model used in Ref.~\cite{stepanov2021} also neglected any depolarization with respect to the pump laser}.
These assumptions are valid if the laser is itself populating the momentum-forbidden dark excitonic reservoir, which is reasonable for the case of a broad pump centered around the exciton energy, as in the experiment.

We estimate the proportionality factor between the laser pump power and the reservoir density using the results for circular polarization, which has the advantage that it only involves $\varepsilon_n$ as a free parameter. 
We determine the LP polaron energy from the peak of the spectral function using the self-energy in Eq.~\eqref{eq:selfenergy} and we then match this theoretical prediction to the reported energy shift~\cite{stepanov2021} of the lower polariton branch at a pump power of 451 pJ. All other parameters (exciton binding energy, Rabi coupling, and detuning) are taken from the experiment~\cite{stepanov2021}. This procedure yields $\varepsilon_n \simeq 2 \Omega_0$ at this pump power; thus, by taking the commonly accepted value for the exciton mass, $m_X= 1.14 m_e$~\cite{kylanpaa2015}, we find a reservoir density of $n \approx 3 \times 10^{12} \text{ cm}^{-2}$, which is comparable to the density reported in the experiment. Note that the fit only depends logarithmically on the scattering energy scale $\varepsilon_X$, and is therefore insensitive to the precise value.

\subsubsection{Reduction of light-matter coupling}

In Fig.~\ref{fig:exptplot}(a-c), we show a comparison between theory and experiment for the lower and upper polariton energies as a function of $\varepsilon_n$~\footnote{The experimental energy shifts were extracted from the observed transmission spectra using a fit with 4 Gaussians, chosen due to the presence of two bare cavity resonances in the spectrum, and due to the relatively high experimental temperature (127 K) involved.}. Importantly, we show here the results for linear polarization, which depend both on the same- and opposite-spin interactions, allowing us to demonstrate the predictive power of our theory. Compared with the circular polarization results above, the theory now depends on one extra parameter, the biexciton binding energy, which was taken to be 20 meV, consistent with various experiments \cite{hao2017,yong2018}. This results in the condition $\epsilon_{LP}\gtrsim-\varepsilon_B$, and we therefore expect the contribution from opposite-spin interactions to be dominant and repulsive. 

We see that the theory predicts the blueshift of the lower polariton in Fig.~\ref{fig:exptplot}(a) very well across a large range of densities. Furthermore, the observed reduction in light-matter coupling (Rabi splitting), due to the substantial blueshift of the lower polariton and minimal blueshift of the upper polariton in Fig.~\ref{fig:exptplot}(b,c), is also quantitatively described by our low-density expressions \eqref{eq:LPupdown} and \eqref{eq:UPupdown}.
We note that including the effects of finite temperature does not substantially alter the quantitative predictions in Fig.~\ref{fig:exptplot}(a-c) as the resulting temperature broadening mostly affects features below the biexciton binding energy. Nevertheless, the reported experimental temperature of 127 K is sufficiently high that the biexciton polariton is unlikely to be visible in this experiment.

\subsubsection{Polarization dependence of the lower polariton blueshift}

We now turn to polarization-dependent measurements. At the same laser power as mentioned above, 451 pJ, the polarization of the incident light was varied continuously, corresponding to changes in the polarizer angle. 

The resulting spectral function is shown in Fig.~\ref{fig:exptplot}(d), together with the experimental transmission spectrum in Fig.~\ref{fig:exptplot}(e). We see that both spectra have a similar structure, with a signal that oscillates as a function of polarization angle, the smallest blueshift occurring for circular polarization ($\theta=0$ mod $\pi/2$) and the largest for linear polarization ($\theta=\pi/4$ mod $\pi/2$). This behavior is highly nontrivial, since the larger linear-polarization blueshift cannot be captured by conventional theories based on the Born approximation~\cite{ciuti1998,tassone1999,glazov2009,Vladimirova2010}, which predict a small $\up\down$ interaction. In the experiment, the large observed blueshift for linear polarization was attributed to an anomalously large saturation; however this could still not explain the fact that the blueshift was largest for linear polarization~\cite{stepanov2021}.

On the other hand, within our theory, the larger linear-polarization blueshift observed in the experiment follows naturally from the resonantly enhanced interactions in the vicinity of the biexciton, which are not present for circular polarization. Furthermore, our calculated spectrum in Fig.~\ref{fig:exptplot}(d) matches the experiment well without needing to introduce any additional fitting. It is thus likely that a dark excitonic reservoir was present in the experiment and dominated the observed interactions.

The full effect of the biexciton resonance is somewhat obscured in this scenario, because the reservoir is copolarized with the polariton. As such, one never has the situation where there are only opposite-spin interactions. In the following section, we shall see how a pump-probe scenario, which allows one to vary the polarization of the reservoir separately from that of the polariton, provides a clear demonstration of the resonantly enhanced interactions. 

\subsection{$\dn$ polariton in a $\up$ exciton medium}

In the second experimental scenario, we consider a pump-probe setup where a spin-$\up$ dark exciton medium is populated by a circularly polarized pump and then probed by a polariton of the opposite polarization ($\down$). This scenario is ideally suited to observing the effects of the biexciton resonance, and it can even potentially be used to probe more exotic few-body resonance phenomena due to triexcitons~\cite{levinsen2019}. Similar pump-probe experiments have been previously realized in semiconductor microcavities~\cite{Fan1998,Saba2000,takemura2014,tan2022}; however these previous works considered the medium to be predominantly polaritonic rather than dark and excitonic. Note that an important distinction between these two cases is that the biexciton resonance in the dark excitonic medium ($\epsilon_{LP} = -\varepsilon_{B}$) 
is significantly shifted from the polariton Feshbach resonance relevant to previous works, where the resonance condition is $\epsilon_{LP} = -\varepsilon_{B}/2$~\cite{Wouters2007,bleu2020}. 

\begin{figure}[tbp] 
\includegraphics[width=1.1\columnwidth, keepaspectratio]{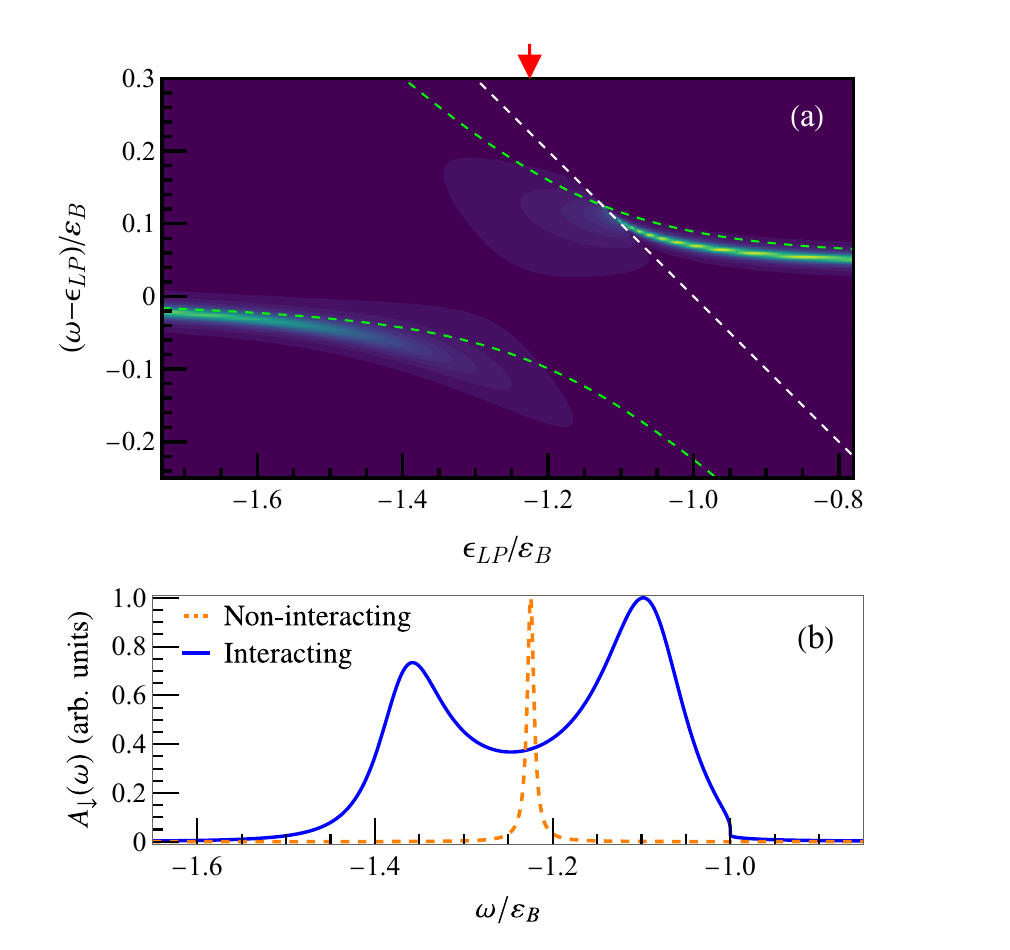}
\vspace{-20pt}
\caption{(a) Spectral function $A_{\dn}(\omega)$ for a spin-$\down$ polariton in a spin-$\up$ incoherent excitonic reservoir. The green dashed lines indicate the energies of the attractive and repulsive polaron branches, calculated using the approximate self-energy as given in Eq.~\eqref{eq:selfenergy}. The white dashed line indicates the biexciton energy, $-\varepsilon_{B}$, below which the reservoir-induced broadening primarily occurs.
(b) Spectral function at $\epsilon_{LP} = -1.225 \varepsilon_{B}$ (indicated by the red arrow in (a)), showing that around resonance both LP and BP polarons can coexist, and that both the existence and broadening of these two features can be attributed to interactions with the dark excitonic medium. Parameters: $\varepsilon_{B} = 20 \text{ meV}$, $\varepsilon_n = 0.17 \varepsilon_{B}$ (corresponding to $n \approx 5 \times 10^{11} \text{ cm}^{-2}$), $k_B T = 0.25 \varepsilon_{B}$, $\Omega_0 = 0.375 \varepsilon_{B}$, $\Gamma_C = 0.005 \varepsilon_{B}$, $\Gamma_X = 0$ \cite{footnotelinewidth}.}
\label{fig:resonance}
\end{figure}

Figure~\ref{fig:resonance} shows our calculated spectral function in the vicinity of the biexciton resonance as a function of bare lower polariton energy $\epsilon_{LP}$, which is equivalent to changing the light-matter detuning. We see that the spectrum displays a significant avoided crossing when the total collision energy matches the energy of the biexciton bound state, i.e., for the resonance condition that occurs in the vicinity of $\omega\simeq-\varepsilon_B$. For the purposes of this illustration, we have used parameters for MoSe$_2$ relevant to the experiment of Ref.~\cite{tan2022}, and we have introduced a reservoir temperature $T\simeq60$K to demonstrate how the incoherence of the reservoir naturally leads to a broadening of the quasiparticles. In particular, this broadening is substantial compared to the case without interactions with the reservoir [Fig~\ref{fig:resonance}(b)], thus providing a natural explanation for why the biexciton feature is often broad in experiment. 
As discussed in Section~\ref{subsec:bxp}, this broadening occurs primarily below the biexciton energy.
It should be noted that our theory also predicts a small temperature-dependent energy shift in the spectrum, which is why the peaks of the spectral function calculated using the approximation in Eq.~\eqref{eq:selfenergy} do not entirely align with those of the finite-temperature spectra in Fig~\ref{fig:resonance}(a).

Qualitatively, our spectra presented in Fig.~\ref{fig:resonance} resemble those observed in \cite{tan2022}. Specifically, Fig.~\ref{fig:resonance}(a) resembles their Fig.~2(d), while Fig.~\ref{fig:resonance}(b) resembles the transmission spectrum shown in Fig.~2(b). However, due to the different resonance condition in the dark reservoir scenario discussed here, we do not need to consider a biexciton that is tighter bound than in previous experiments on monolayer MoSe$_2$. This is unlike the experiment~\cite{tan2022}, which took the biexciton binding energy to be 29 meV, which is even larger than previous reports of the trion binding energy (around 25 meV)~\cite{tan2020}. We stress that the experiment is likely to contain both polaritons as well as a dark excitonic reservoir, and therefore a quantitative comparison with experiment would need to account for both contributions and is beyond the scope of the present work.

\section{Conclusions}
\label{sec:conclusion}

In conclusion, we have developed a theory of exciton polaritons in an incoherent excitonic reservoir which exposes the polaronic nature of the light-matter coupled quasipartcles. By incorporating the exact low-energy interactions with the medium into a many-body Green's function formalism, we have demonstrated that interactions with the reservoir lead to a wealth of intriguing effects, many of which are missed by standard mean-field approaches. Our predictions include a medium-induced reduction of the light-matter coupling and Rabi splitting, without needing to invoke Pauli blocking, and the appearance of a new quasiparticle, the biexciton polariton polaron. We expect the physics that we have uncovered here to apply to polariton systems more generally, thus providing a new understanding of the physical mechanism responsible for the saturation of the Rabi splitting.

The presence of an excitonic reservoir is already likely to play a non-trivial role in current experiments, particularly those with a non-resonant pump where the reservoir density may be significant, or those where the lower polariton is close to the biexciton resonance such as is generally the case in monolayer TMD microcavities. Our simple analytic expressions for the modifications of energies, exciton and photon fractions, and lifetimes are likely to aid future experiments in ascertaining the impact of the reservoir on the observed physics.

We have furthermore analyzed two particular experimentally relevant scenarios, related to recent experiments carried out in MoSe$_2$ monolayer microcavities~\cite{stepanov2021,tan2022}. For the case of a polariton co-polarized with the reservoir, we have demonstrated that the blueshift of the lower polariton can be largest for linear polarization due to the presence of the biexciton resonance. Our results are in quantitative agreement with those of Ref.~\cite{stepanov2021} with only a single fitting parameter, the reservoir density, which is particularly remarkable given that the polarization-dependent behavior of the blueshift could not even be qualitatively explained by conventional mean-field theories~\cite{stepanov2021}. In the case of a counter-polarized reservoir, we have shown that the incoherent nature of the reservoir can lead to a significant broadening of the biexciton feature, in agreement with the observations of Ref.~\cite{tan2022} as well as previous measurements~\cite{Saba2000,takemura2014}. Thus, the significant biexciton linewidth observed in experiment may arise from the tail of the exciton momentum distribution, rather than due to an intrinsic property of the biexciton.

Our results open up new possibilities to engineer and manipulate exciton-polaritons as well as their interactions. For example, they have immediate implications for optical trapping of polariton condensates, a key experimental technique due to its great flexibility \cite{sanvitto2011, Tosi2012}. Recently, there have been experimental studies on `reservoir optics', i.e., non-resonantly pumping the system across a spatial area, and using the resulting reservoir to focus and trap polariton condensates \cite{schmutzler2015, wang2021, Aristov2023}. Our work suggests that since the polariton-reservoir interaction can be dramatically enhanced and even its sign reversed, these reservoir optics devices can be engineered to provide strong responses through the manipulation of the laser spin and photon-exciton detuning, while at the same time maintaining the flexibility and predictability of systems in which the laser alone provides a well-understood trapping potential. In particular, this potentially provides a versatile platform for achieving strongly correlated photons via reservoir engineering of the polariton polarons' mutual interactions~\cite{vishnevsky2014}.

\acknowledgments 
We acknowledge useful discussions with Dmitry Efimkin, Brendan Mulkerin, Maxime Richard, Amit Vashisht, Anna Minguzzi, Francesca Maria Marchetti, Antonio Tiene, and Maciej Pieczarka. 
We also thank Maxime Richard and Amit Vashisht for sharing the experimental data of Ref.~\cite{stepanov2021}. We acknowledge support from the Australian Research Council Centre of Excellence in Future Low-Energy Electronics Technologies (CE170100039). 
JL and MMP are also supported through the Australian Research Council Future Fellowships FT160100244 and FT200100619, respectively. 
KC acknowledges support from an Australian Government Research Training Program (RTP) Scholarship.
\input{appendix.tex}
\bibliography{resbib}

\end{document}

%% file: appendix.tex
\appendix

\section{Equation of motion and Green's functions in the absence of interactions}
\label{app:eom}
In this appendix, we consider the light-matter coupled system in the absence of exciton-exciton interactions. In this case, the operators evolve according to the non-interacting Hamiltonian (i.e., $\hat{c}_{\sigma}(t) = e^{i \hat{H}_0 t} \hat{c}_{\sigma} e^{-i \hat{H}_0 t}$). The photon and $\k=0$ exciton Green's functions take the form
\begin{subequations}
\begin{align}
\label{eq:gctimedefapp}
    G^{(0)}_{C}(t) &= -i \theta(t) \braket{\hat{c}_{\sigma}(t) \hat{c}^{\dagger}_{\sigma}(0)} ,\\ 
G^{(0)}_{X}(t) &= -i \theta(t) \braket{\hat{x}_{0 \sigma}(t) \hat{x}^{\dagger}_{0 \sigma}(0)}, \label{eq:gxtimedef}
\end{align}
\end{subequations}
where, for simplicity, we drop the spin labels on the Green's functions since these are independent of spin in the absence of interactions. We also have the two off-diagonal Green's functions,
\begin{subequations}
\begin{align}
    G^{(0)}_{XC}(t) &= -i \theta(t) \braket{\hat{x}_{0\sigma}(t) \hat{c}^{\dagger}_{\sigma}(0)}, \\
    G^{(0)}_{CX}(t) &= -i \theta(t) \braket{\hat{c}_{\sigma}(t) \hat{x}^{\dagger}_{0 \sigma}(0)}.
\end{align}
\end{subequations}
These are conveniently arranged into a matrix in the exciton-photon basis,
\begin{equation}
\mathbf{G}^{(0)}(t) = 
\begin{pmatrix}[1.6] 
G^{(0)}_{X}(t) & G^{(0)}_{XC}(t)
\\ G^{(0)}_{CX}(t) & G^{(0)}_{C}(t) \end{pmatrix}.
\label{eq:greenmatdef}
\end{equation}

The Heisenberg equations of motion $i \partial_t \hat{x}_{0 \sigma}(t) = [\hat{x}_{0 \sigma}(t), \hat{H}_0]$ and $i \partial_t \hat{c}_{\sigma}(t) = [\hat{c}_{\sigma}(t), \hat{H}_0]$ yield
\begin{subequations}
\begin{align}
    i \partial_t \hat{x}_{0 \sigma}(t)&=\Omega_0 \hat{c}_{\sigma}(t), \\
    i \partial_t \hat{c}_{\sigma}(t)&=\Delta_0 \hat{c}_{\sigma}(t) + \Omega_0 \hat{x}_{0 \sigma}(t),
\end{align}
\end{subequations}
respectively. Then, combining the equations of motion with the definitions of the Green's functions gives
\begin{subequations}
    \begin{align}
    i \partial_t G^{(0)}_{C} (t) &= \delta(t) 
    + \Delta_0 G^{(0)}_{C}(t) + \Omega_0 G^{(0)}_{XC}(t), \\
    i \partial_t G^{(0)}_{X} (t) &= \delta(t) 
    + \Omega_0 G^{(0)}_{CX}(t),\\
    i \partial_t G^{(0)}_{XC} (t) &= 
    \Omega_0 G^{(0)}_{C}(t),\\
    i \partial_t G^{(0)}_{CX} (t) &= 
    \Delta_0 G^{(0)}_{CX}(t) + \Omega_0 G^{(0)}_{X}(t).
    \end{align}
\end{subequations}
Here, the Dirac delta function $\delta(t)$ comes from taking the derivative of the Heaviside function $\Theta(t)$, and we have used the fact that $\expval*{\hat{x}_\sigma\hat{c}_\sigma^\dag}=0$ and $\expval*{\hat{x}_\sigma\hat{x}_\sigma^\dag} = \expval*{\hat{c}_\sigma\hat{c}_\sigma^\dag} =1$.
Fourier transforming, this set of equations yields an explicit result for Eq.~\eqref{eq:greenmatdef},
\begin{equation}
\mathbf{G}^{(0)}(\omega) = \frac{1}{\omega(\omega - \Delta_0)-\Omega_0^2}
\begin{pmatrix} 
\omega-\Delta_0 & \Omega_0
\\ \Omega_0 & \omega \end{pmatrix}.
\end{equation}
This is Eq.~\eqref{eq:G0matrix} of the main text.

\onecolumngrid
\section{Two-body \textit{T} matrix}
\label{app:Tmat}

The two-body $T$ matrix, $\mathcal{T}_{\sigma\sigma'}(\p_1,\p_2;\Q,\omega)$, corresponds to the sum of all scattering processes of two particles, where particle 1 has spin $\sigma$ and incoming (outgoing) momentum $\Q/2-\p_1$ ($\Q/2-\p_2/2$), particle 2 has spin $\sigma'$ and incoming (outgoing) momentum $\Q/2+\p_1$ ($\Q/2+\p_2/2$), while $\omega$ is the total frequency. This satisfies the relation depicted in Fig.~\ref{fig:diagrams}(c):
\begin{subequations}
\label{eq:Tmatapp}
\begin{align}
    \mathcal{T}_{\sigma\sigma'}(\p_1,\p_2;\Q,\omega) 
    & = v_{\sigma\sigma'}(\p_1-\p_2)+\sum_{\P} v_{\sigma\sigma'}(\p_1-\P)\frac{1}{\omega-\epsilon_{-\P+\Q/2}-\epsilon_{\P+\Q/2}}v_{\sigma\sigma'}(\P-\p_2)+\dots \\
    & = v_{\sigma\sigma'}(\p_1-\p_2)+\sum_{\P} v_{\sigma\sigma'}(\p_1-\P)\frac{1}{\omega-\epsilon_{-\P+\Q/2}-\epsilon_{\P+\Q/2}}\mathcal{T}_{\sigma\sigma'}(\p_1,\P;\Q,\omega).
\end{align}
\end{subequations}
To make connection to Eq.~\eqref{eq:gammap}, we first note that the solution of that equation separates into two parts, with the exchange contribution being proportional to $\delta_{\sigma\sigma'}$. We then recognize that the momenta in Eq.~\eqref{eq:gammap} correspond to the particular choices $\P=\Q/2-\k$, $\p_1=\Q/2-\p$ and $\p_2=\pm\Q/2$, with the plus and minus signs corresponding to the direct and exchange terms, respectively. Finally, upon formally expanding Eq.~\eqref{eq:gammap} similarly to Eq.~\eqref{eq:Tmatapp}, we note that the series contains precisely the same terms as Eq.~\eqref{eq:Tmatapp}, apart from the leading interaction $v_{\sigma\sigma'}$ ($G_{X\sigma}$ just acts as a multiplicative prefactor). Taking $\omega\to\omega+\epsilon_\Q$ and $\p=0$ we then finally arrive at Eq.~\eqref{eq:Gamma}.

\section{Derivation of the polaron energies at low-densities}
\label{app:applowdens}

In the limit of a small reservoir density and a minimal occupation of excitons for which $\epsilon_\Q>\Omega_0$ (such as at low temperature), we can derive perturbatively exact results for the energy shifts of the polariton branches. Let us first consider the shift of the lower and upper polariton branches, assuming that the lower polariton is not resonant with the biexciton. At low density, we can use the self-energy~\eqref{eq:selfenergy} which, since the energy is approximately $\epsilon_{LP,UP}$, takes the form
\begin{align}
    \Sigma_{\sigma}(\omega) &\simeq \Sigma_{\sigma}(\epsilon_{LP, UP}) = \frac{2 \varepsilon_{n_{\sigma}}}{\ln{(-\varepsilon_{X}/\epsilon_{LP, UP})}} + \frac{\varepsilon_{n_{\bar\sigma}}}{\ln{(-\varepsilon_{B}/\epsilon_{LP, UP})}}.
\end{align}
To obtain the quasiparticle energies, we then apply the condition $\mathrm{Re}[G_{C\sigma}^{-1}(\omega)]=0$. Since the self-energy is now approximately a constant, it simply provides a rigid shift of the exciton energy. Furthermore, the imaginary part of the self-energy only enters into the quasiparticle energies at higher order in the density. Therefore
\begin{align}
    \epsilon_{LP, UP\sigma}(n) &\simeq \frac{1}{2} \big( \Delta_0 + \mathrm{Re}[\Sigma_{\sigma}(\epsilon_{LP, UP})] \pm \sqrt{(\Delta_0 - \mathrm{Re}[\Sigma_{\sigma}(\epsilon_{LP, UP})])^2 + 4\Omega_0^2} \big) \nn\\
    \label{eq:qpslowdens}
    &\simeq \epsilon_{LP, UP}+\frac{1}{2} \mathrm{Re}[\Sigma_{\sigma}(\epsilon_{LP, UP})] \left( 1 \pm \frac{\Delta_0}{\sqrt{\Delta_0^2 + 4\Omega_0^2}}\right)\nn\\ & = \left\{\begin{array}{ll}
    \epsilon_{LP}+|X_0|^2 \Sigma_{\sigma}(\epsilon_{LP}) & \mathrm{LP} \\
    \epsilon_{UP}+|C_0|^2 \mathrm{Re}[\Sigma_{\sigma}(\epsilon_{UP})] & \mathrm{UP} \end{array}\right.,
\end{align}
where the upper/lower signs correspond to the UP/LP. In the last step we recognized the exciton fractions of the lower and upper polaritons, $|X_0|^2$ and $|C_0|^2$, respectively, see Eq.~\eqref{eq:hopfield}. Upon explicit evaluation of the self-energy, Eq.~\eqref{eq:qpslowdens} yields the expressions Eq.~\eqref{eq:LPupdown} and \eqref{eq:UPupdown}.

In the case of the biexciton polariton polaron, we cannot follow the same procedure and simply substitute the value of the biexciton energy for \(\omega\), as the logarithm in the self-energy diverges at this point. Instead, we perform a pole expansion around this point, i.e., we replace \(\omega \rightarrow -\varepsilon_{B} + \eta\) for some small parameter \(\eta\). 
In the vicinity of the resonance, the contribution from interactions with same-spin excitons can be ignored as the resonant contribution from the biexciton is far greater. From the equation for the quasiparticle energies, $\mathrm{Re}[G_C^{-1}(\omega_p)]=0$, we then find an equation for $\eta$
\begin{equation}
    \left(-\varepsilon_{B} + \eta - \Sigma_{\sigma}(-\varepsilon_{B} + \eta)\right)(-\varepsilon_{B} + \eta - \Delta_0) - \Omega_0^2 = 0.
    \label{eq:poles}
\end{equation}
Using the fact that to leading order, we have
\begin{equation}
    \Sigma_{\sigma}(-\varepsilon_{B} + \eta) \simeq \frac{\varepsilon_{B} \varepsilon_{n_{\bar\sigma}}}{\eta},
\end{equation}
one can expand Eq.~\eqref{eq:poles} and keep the leading order term in $\eta$, we obtain
\begin{equation}
-\Omega_0^2 + (\Delta_0 + \varepsilon_{B})\varepsilon_{B}
+ \varepsilon_{n_{\bar\sigma}}\varepsilon_{B}\frac{\Delta_0 + \varepsilon_{B}}{\eta} = 0.
\end{equation}
Solving for this equation for $\eta$ and reminding that we have $\omega=\eta-\epsilon_B$ we obtain the BP energy
\begin{align}
     \epsilon_{BP \sigma}(n) \simeq -\varepsilon_{B}\left(1+  \frac{\varepsilon_{n_{\bar\sigma}}}{\varepsilon_{B} - \frac{\Omega_0^2}{\Delta_0 + \varepsilon_{B}}}
     \right),
\end{align}
which is Eq.~\eqref{eq:biexcFeature} of the main text.

Finally, we can also obtain perturbative results for the energies when the lower polariton energy is tuned to be resonant with the biexciton, i.e., when $\epsilon_{LP}\simeq -\varepsilon_B$. Here again, we have to do a pole expansion near \(\omega \simeq -\varepsilon_{B} \), but this time, we also impose the resonance condition \(\epsilon_{LP}=-\varepsilon_{B}\), which requires us to keep higher order terms in $\eta$ when approximating the self energy
\begin{equation}
    \Sigma_{\sigma}(-\varepsilon_{B} + \eta) \simeq \frac{\varepsilon_{n_{\bar\sigma}}}{\frac{\eta}{\varepsilon_{B}} + \frac{\eta^2}{ 2 \varepsilon_{B}^2
    } }.
\end{equation}
Substituting this expression into Eq.~\eqref{eq:poles}, and keeping the leading order terms in $\eta$ one obtains the quadratic equation 
\begin{equation}
  \left(\frac{\Delta_0}{\varepsilon_B} +2\right)\eta^2 + \varepsilon_{n_{\bar\sigma}} \eta - \varepsilon_{n_{\bar\sigma}} (\varepsilon_B+\Delta_0) =0.
\end{equation}
To the leading order in $\varepsilon_{n_{\bar\sigma}}$, the two solutions of this equation allows us to find 
\begin{align}
    \epsilon_{\pm, \sigma}(n) &\simeq 
    -\varepsilon_B\left(1 \pm \sqrt{\frac{\varepsilon_{n_{\bar\sigma}}}{\varepsilon_B}\frac{1+\Delta_0/\varepsilon_B}{2+\Delta_0/ \varepsilon_{B}}}\right),
\end{align}
which is Eq.~\eqref{eq:resonance} of the main text.

\twocolumngrid

%% file: main.bbl
\begin{thebibliography}{101}
\bibitem [{foo()}]{footnotelinewidth}%
  \BibitemOpen
  \href@noop {} {}\bibinfo {note} {The exciton and photon linewidths $\Gamma_X$ and $\Gamma_C$ were included in the photonic Green's function as follows: $G_{C \sigma}(\omega) = (\omega - \Delta_0 + i \Gamma_{C} - \frac{\Omega^2_0}{\omega + i \Gamma_{X} - \Sigma_{\sigma}(\omega)})^{-1}$}\BibitemShut {NoStop}%
\bibitem [{ste()}]{stepanovfootnote}%
  \BibitemOpen
  \href@noop {} {}\bibinfo {note} {Uncertainties of peak positons were given by the standard error of a 4-Gaussian fit to the observed transmission spectrum, and then propagated in quadrature.}\BibitemShut {Stop}%
\end{thebibliography}
